\documentstyle[epsfig,aps,prb,twocolumn,floats,eqsecnum]{revtex}

\begin{document}

\textheight23.7cm

\newcommand{\gsim}{
\,\raisebox{0.35ex}{$>$}
\hspace{-1.7ex}\raisebox{-0.65ex}{$\sim$}\,
}

\newcommand{\lsim}{
\,\raisebox{0.35ex}{$<$}
\hspace{-1.7ex}\raisebox{-0.65ex}{$\sim$}\,
}

\bibliographystyle{prsty}

\title{ \begin{flushleft}
{\small 
PHYSICAL REVIEW B $\qquad$
\hfill
VOLUME {\normalsize 56}, 
NUMBER {\normalsize 17}
\hfill 
{\normalsize 1} NOVEMBER {\normalsize 1997-}I, {\normalsize 11102$-$11118} 
}\\
\end{flushleft}  
Thermally activated resonant magnetization tunneling in molecular magnets: \\ Mn$_{12}$Ac and others
}

\author{
D.~A. Garanin
\renewcommand{\thefootnote}{\fnsymbol{footnote}}
\footnotemark[1]
and E.~M. Chudnovsky
\renewcommand{\thefootnote}{\fnsymbol{footnote}}
\footnotemark[2]
}

\address{
Department of Physics and Astronomy, City University of New York, 
Lehman College,\\
Bedford Park Boulevard West, Bronx, New York 10468-1589 \\
\smallskip
{\rm (Received 27 May 1997)}
\bigskip\\
\parbox{14.2cm}
{\rm
The dynamical theory of thermally activated resonant magnetization
tunneling in uniaxially anisotropic magnetic molecules 
such as Mn$_{12}$Ac ($S=10$) is developed.
The observed slow dynamics of the system is described by master equations
for the populations of spin levels.
The latter are obtained by the adiabatic elimination 
of fast degrees of freedom from the density matrix equation
with the help of the perturbation theory developed earlier for the tunneling
level splitting [D. A. Garanin, J. Phys. A, {\bf 24}, L61 (1991)].
There exists a temperature range (thermally activated tunneling) where the
escape rate follows the Arrhenius law, but has a nonmonotonic dependence on
the bias field due to tunneling at the top of the barrier.
At lower temperatures this regime crosses over to the non-Arrhenius law
(thermally assisted tunneling).
The transition between the two regimes can be first or second order, 
depending on the transverse field, which can be tested in experiments.
In both regimes the resonant maxima of the rate occur when spin levels in the
two potential wells match at certain field values.
In the thermally activated regime at low dissipation each resonance has a multitower self-similar structure
with progressively narrowing peaks mounting on top of each other. 
[S0163-1829(97)00141-0]
\smallskip
\begin{flushleft}
PACS number(s): 75.45.+j, 75.50.Tt
\end{flushleft}
} 
} 
\maketitle

\section{Introduction}

In recent years there has been great experimental and theoretical effort to observe and interpret quantum tunneling of magnetization in monodomain particles.
The interest in this problem arises from the fact that the magnetization ${\bf M}$ of a particle containing a few thousand atoms is a macroscopic degree of freedom.
Thus tunneling of the particle's magnetization between different equilibrium orientations at low temperatures requires strong coherence between  atomic spins and may be very sensitive to the interaction with the environment.
A similar problem has been extensively studied in superconductors in the context of macroscopic quantum tunneling, where good agreement has been achieved between theory \cite{calleg83} and experiment.  \cite{claetal88}
Observation of magnetization tunneling is complicated by the difficulty in preparing identical magnetic particles.
Experiments have been performed \cite{tejzha95} on particles distributed over sizes and shapes.
These experiments revealed temperature-independent magnetic relaxation which was attributed to tunneling.
When an effort was made to narrow the distribution, resonance was observed  \cite{awsetal92,gidetal95} in the absorption of the ac field, similar to the tunneling resonance in the ammonia molecule.

Difficulties in manufacturing identical magnetic particles for tunneling experiments have led to new techniques of measuring individual particles \cite{weretal96,weretal97} and to
the idea of searching for magnetization tunneling in magnetic molecules of large spin.
The system that caught the most recent attention is the crystal 
Mn$_{12}$ acetate (Mn$_{12}$Ac) having the chemical formula
[Mn$_{12}$O$_{12}$(CH$_3$COO)$_{16}$(H$_2$O)$_4$]$\cdot$2CH$_3$COOH$\cdot$4H$_2$O.
This compound has been synthesized by Lis,\cite{lis80} but its physical properties had not received much attention until Sessoli {\em et al.} \cite{sesgatcannov93} noticed magnetic bistability of this system.
In the Mn$_{12}$Ac molecule the 12 Mn ions are strongly bound ferrimagnetically via the superexchange through oxygen bridges.
These molecules behave effectively as magnetic clusters of spin $S=10$,
\cite{sesgatcannov93} as has been confirmed
by the Curie-law temperature dependence of the susceptibility $\chi$.
As follows from the very low value of the Curie constant, 
$\Theta_C\approx -0.05$ K, \cite{novses95} the interaction between the
Mn$_{12}$Ac molecules is very weak, 
presumably of the dipole-dipole origin.
Mn$_{12}$Ac is characterized by a very strong uniaxial anisotropy
${\cal H}_{\rm A}=-DS_z^2$, where $D\simeq 0.72$ K from high-field
EPR, \cite{canetal91} $D\simeq 0.75$ K from single-crystal magnetic
susceptibility \cite{ses95} measurements, and $D\simeq 0.77$ K 
from neutron scattering experiments. 
\cite{henetal97}
This leads to a barrier of about $U=DS^2\simeq 75$ K between the states 
$\pm S$.
Note, however, that experiments on resonant spin tunneling 
\cite{frisartejzio96} (see below) suggest a value of $D$ close 
to 0.6 K and correspondingly the barrier height of 60 K.

The advantage of Mn$_{12}$Ac and other molecular magnets is that they 
are rather simple model systems, which facilitates their theoretical
consideration and interpretation of experiments.
Of course, it should be understood that a cluster of spin 10 cannot
be treated macroscopically.
The limit of macroscopic quantum tunneling is the one where the
quantization of spin levels is irrelevant.
On the contrary, in the Mn$_{12}$Ac cluster the distance between the
ground-state and the first excited level is 12--15 K.
At low temperature quantization of levels must, therefore, 
dominate the properties of the system. 
In this sense Mn$_{12}$Ac is closer to conventional quantum-mechanical
systems where tunneling is of a resonant character.
Nevertheless, as we shall see, the high value of spin leads to the 
macroscopic time scale for the dynamics of the magnetization, which has
been tested in macroscopic experiments.

An important feature of Mn$_{12}$Ac is that if no strong transverse 
field is applied to the system, the interactions responsible for 
tunneling are small in comparison to the anisotropy energy 
${\cal H}_{\rm A}=-DS_z^2$ which itself conserves the $S_z$ component
of the spin.
As a result of this and of the large spin of the system, the tunneling 
between low-lying energy levels should be extraordinary slow, which 
makes Mn$_{12}$Ac an excellent candidate for information storage at
the molecular level.
Another possible application of molecular magnets is that for 
quantum computing.
For that application tunneling between the low-lying states should be 
made more pronounced, and the interaction with the environment 
destroying coherent oscillations of the spin between two wells should 
be kept small.
This is hardly the case for Mn$_{12}$Ac where nuclear spins of 
manganese atoms strongly suppress the coherence. \cite{garg95nuccoh}
The example of Mn$_{12}$Ac is, however, instructive since other
systems with similar properties can be developed, which could be better
candidates for quantum computation.

The first indications of magnetization tunneling in Mn$_{12}$Ac 
were seen in the magnetization relaxation experiments of Paulsen and 
Park \cite{paupark95} and the dynamic susceptibility measurements of 
Novak and Sessoli. \cite{novses95}
The measured relaxation rate of Mn$_{12}$Ac followed the Arrhenius law 
$\Gamma=\Gamma_0\exp(-U_{\rm eff}/T)$ 
with the peaks at some values of the longitudinal field $H_z$.
These peaks were interpreted \cite{novses95} as the resonant 
thermally assisted tunneling between the levels near the  top of the barrier , 
which decreased the effective barrier height  $U_{\rm eff}$.
Subsequent dynamic hysteresis experiments \cite{frisartejzio96} have 
proved that conjecture as they have shown many regularly spaced steps 
in the hysteresis loop at the values of $H_z$ at which the levels on 
both sides of the barrier come into resonance 
(see also Refs. \onlinecite{heretal96,thoetal96,heretal97,lioetal97}).
These steps indicate an increased relaxation rate at the 
corresponding bias fields $H_z$.
Very recently a similar observation was made on Mn$_{12}$ phosphat 
\cite{aubetal97} which was described as a magnetic cluster of spin 
$S=9.5$.

The transverse field $H_x$ applied to a uniaxial magnetic system mixes 
the unperturbed energy levels and enhances tunneling.
The search for an increased tunneling in the transverse field has been
undertaken in recent hysteresis \cite{frietal97} and dynamic susceptibility 
\cite{luisetal97} measurements. 
The results show that the speeding up of the relaxation can be 
explained mostly through the classical effect of the barrier lowering 
in a transverse field, whereas the resonant tunneling peaks remaining 
after subtraction of this main effect are nearly independent of $H_x$.
Actually both effects come from the same source: The classical height of
the barrier can be determined quantum mechanically from the condition that the tunneling level splitting becomes comparable with the level spacing, 
which means strong {\em nonresonant} tunneling, i.e., the absense
of a barrier at that level. \cite{chufri}

A large number of experimental observations of magnetization 
tunneling in molecular magnets has been accumulated to date and the major 
relevant physical processes have been identified.
A theoretical framework for the dynamical description of the 
combined process of the thermal activationand tunneling in these 
materials is still lacking, however.
In particular, the form and the width of the tunneling peaks measured 
in experiments has not yet been explained.
The aim of this article is to supply an appropriate theory.

The idea of the work is to apply the density matrix formalism in the 
case when the tunneling is caused by a transverse field $H_x$ which is
small enough and can be considered as a perturbation.
The applicability criterium of this method is 
$H_x \ll H_{\rm A}$, where 
$H_{\rm A} \equiv (2S-1)D/(g\mu_B)$ is the anisotropy field.
The latter in turn coincides with the critical value of the transverse
field at which in the classical case of $S\gg 1$ the double-well 
structure of the spin energy disappears.
For Mn$_{12}$Ac, the anisotropy field is of order 10 T, so that the condition
$H_x \ll H_{\rm A}$ allows for rather large $H_x$.
In this relevant range of the transverse field one can use the 
physically transparent and technically convenient basis of the 
eigenfunctions of the anisotropy energy ${\cal H}_{\rm A}=-DS_z^2$.
The slow dynamics of the system driven by the thermal activation and 
tunneling processes can be described with the help of the adiabatic 
elimination of the fast degrees of freedom in the density matrix.
The latter is a dynamical generalization of the calculation of the 
tunneling level splittings
in the high orders of the perturbation theory. \cite{gar91jpa}

The remaining part of the paper is organized as follows.
In Sec. \ref{qmech}
the properties of an {\em isolated} magnetic cluster in a transverse 
field are briefly reviewed and the perturbation theory is compared 
with other approaches to the problem.
In Sec. \ref{dme}
the density matrix equation (DME) for the uniaxial magnetic system 
interacting with a phonon bath is formulated and discussed.
In Sec. \ref{slowdyn}
the fast degrees of freedom in the DME are eliminated and a simplified
system of equations describing the slow spin dynamics in terms of the 
diagonal and antidiagonal matrix elements connecting resonant pairs 
of levels in different wells is derived.
It is shown that the level broadening due to the interaction with the 
environment suppresses coherent oscillations and, if strong enough, 
makes the motion of the spin between two degenerate levels overdamped.
In this case, and also in the case of the thermally assisted quantum 
tunneling, when the relaxation rate is limited by the exponentially 
slow process of climbing up the energy barrier, the DME further 
simplifies to the system of kinetic balance equations for the level 
populations $N_m$ only.
The latter describes the hopping of particles between adjaicent energy
levels and through the barrier.
In Sec. \ref{kram}
the system of equations for the level populations $N_m$ is solved 
analytically in the Arrhenius regime $T\ll U \approx DS^2$.
In Sec. \ref{tat}
the transition from the Arrhenius regime to pure quantum tunneling
at lower temperatures is discussed. 
In Sec. \ref{numres}
the numerical results for the dependences of the escape rate on
longitudinal and transverse fields in the Arrhenius regime are 
presented.
Here we also analyze the influence of the Mn nuclear spins
and a small scatter of the easy-axis directions in the oriented
polycrystalls on resonant magnetization tunneling.
In Sec. \ref{disc}  
further developments of the theory and suggestions for experiments are
discussed.

\section{Tunneling level splitting and classical barrier lowering}
\label{qmech}

\begin{figure}[t]
\unitlength1cm
\begin{picture}(11,10.5)
\centerline{\epsfig{file=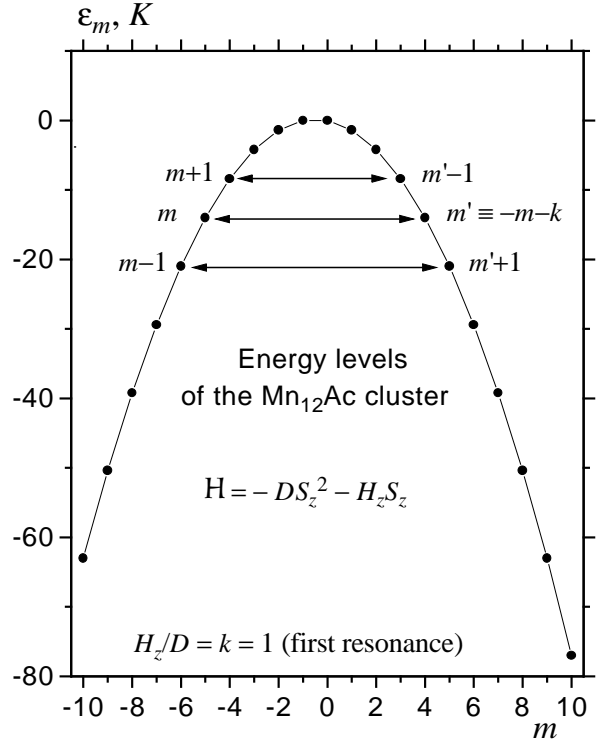,angle=0,width=10cm}}
\end{picture}
\caption{ \label{t_levs} 
Spin energy levels of a Mn$_{12}$Ac molecule for $H_x=0$ and 
$H_z = D$ corresponding to the first resonance, $k=1$, in
Eq. (\protect\ref{hzres}). 
}
\end{figure}

The spin Hamiltonian of an isolated Mn$_{12}$Ac molecule in magnetic field 
${\bf H}$ can be written in the form
\begin{equation}\label{spinham}
{\cal H} = -D S_z^2 - H_z S_z - H_x S_x,
\end{equation}
where ${\bf H}$ stands for $g\mu_B{\bf H}$ with $g\simeq 1.9$.
Henceforth we will usually drop the combination $g\mu_B$ for better 
readability of the formulas.
The system is described by the $2S+1$ energy levels which in the absense of the 
transverse field $H_x$ are labeled by the spin projection $m$ on the $z$ axis 
and given by 
$\varepsilon_m = -Dm^2 - H_z m$
(see Fig. \ref{t_levs}).
It can be easily checked that for the regularly spaced values of the 
longitudinal field $H_z$ satisfying
\begin{equation}\label{hzres}
H_z = H_{zk} = kD, \qquad k=0, \pm 1, \pm 2, \ldots,
\end{equation}
the energy levels on both sides of the barrier are pairwise degenerate
\begin{equation}\label{epsres}
\varepsilon_m = \varepsilon_{m'}, \qquad m<0, \qquad m' = - m - k . 
\end{equation}

The latest high-field EPR experiments \cite{bargatses97} suggest that
there are correction terms of the types $-AS_z^4$ and 
$-B(S_+^4 + S_-^4)$
in the spin Hamiltonian (\ref{spinham}) of Mn$_{12}$Ac.
This means that the degeneracy of different level pairs $m,m'$
is actually achieved at slightly different values of $H_z$.
We shall, however, ignore this effect in the following since it
does not significantly change the results.
As we shall see, only one or maximally two pairs of degenerate
levels contribute to resonant tunneling, and hence the lack of 
simultaneous degeneracy of all appropriate level pairs is unimportant.

The model Hamiltonian (\ref{spinham}) was a whetstone for
different theories of spin tunneling long before its relevance
for Mn$_{12}$Ac and other molecular magnets had been established.
In the quasiclassical limit $S\gg 1$, the rate of tunneling from
the ground-state for different values of $H_z$ was calculated by
Chudnovsky and Gunther \cite{chugun88} with an exponential accuracy
with the help of the instanton technique.
Enz and Schilling \cite{enzsch86} developed a more sophisticated version
of the
instanton approach to spins to obtain the ground-state tunneling level 
splitting with the prefactor.
The latter result was rederived by Zaslavskii \cite{zas90} by a more simple
method based on the mapping onto a particle problem.
Also, van Hemmen and S\"ut{\H o} \cite{lvhsuto86epl,lvhsuto86} formulated
the WKB method for spin systems and calculated the tunneling rates and
corresponding level splittings for the excited states of Eq. (\ref{spinham}).
Scharf, Wreszinski, and van Hemmen \cite{schwrelvh87} proposed an 
approach based on a particle mapping with subsequent application of the
WKB approximation to refine the results for the 
splittings of excited levels for systems with moderate spin.
The applicability of this approach is confined, however, to the limit
of small transverse fields $H_x$, where it is still possible to label 
the energy levels of Eq. (\ref{spinham}) by the quantum number $m$.

In the case of small $H_x$, which, as we shall see, is relevant
for magnetic clusters with moderate spin, the level splittings
can be calculated in a more direct and simple way using the
high-order perturbation theory.
An early application of this method is due to Korenblit and
Shender \cite{korshe78rus} who studied ground-state splitting in
rare-earth compounds having high spin values (e.g., $S=8$ for Ho).
Garanin \cite{gar91jpa} has derived a formula for the splitting of all
levels of the Hamiltonian (\ref{spinham}).
A recent remake of the method is due to Hartmann-Boutron. \cite{harbou95} 
Schatzer, Breymann, and Thomas \cite{schbretho96} extended the 
perturbative approach to describe tunneling in a system of two spins.

In the general biased case, the tunneling level splitting of the 
resonant level pair $m,m'$ appears, minimally, in the $|m-m'|$th order 
of a perturbation theory and is given by the shortest chain of matrix 
elements and energy denominators connecting the states $m$ and $m'$, 
\begin{eqnarray}\label{chain}
&&
\Delta \varepsilon_{mm'} = 2 
V_{m,m+1} \frac{1}{ \varepsilon_{m+1} - \varepsilon_m } 
\nonumber\\
&&
\qquad \times V_{m+1,m+2} \frac{1}{ \varepsilon_{m+2} - \varepsilon_m } \cdots
V_{m'-1,m'},
\end{eqnarray}
where  
\begin{equation}\label{lm}
V_{m,m+1} = \langle m | H_x S_x | m+1 \rangle = 
\frac 12 H_x l_{m,m+1},
\end{equation}
$l_{m,m+1} \equiv \sqrt{S(S+1)-m(m+1)}$ are the matrix elements of the operator $S_x$, which are symmetric functions of their arguments, and 
$\varepsilon_m = -Dm^2 - H_z m$ are the unperturbed energy
levels.  
The calculation in Eq. (\ref{chain}) for the arbitrary resonance number $k$
yields the formula \cite{chufri}
\begin{eqnarray}\label{split}
&&
\Delta \varepsilon_{mm'} = \frac{2D}{[(m'-m-1)!]^2}
\nonumber\\
&&
\qquad
\times \sqrt{\frac{ (S+m')! (S-m)! }{ (S-m')! (S+m)! } }
\left( \frac{ H_x }{ 2D } \right)^{m'-m},
\end{eqnarray}
which is the generalization of the zero-bias result of 
Ref. \onlinecite{gar91jpa}.
Note that here, according to the convention of Eq. (\ref{epsres}), $m<0$,
$m'\equiv -m-k$, and hence $m'-m>0$.
Equation (\ref{chain}) describes the interaction between the pair of resonant
levels $m,m'$ through the intermediate levels in the virtual state.
As is well known for the two-state problem, the splitting 
$\Delta \varepsilon_{mm'}$ is exactly equal to the tunneling frequency
$\Omega_{mm'}$ with which the probability of finding the system in one
of these states oscillates with time if the initial condition is an
unperturbed eigenstate. 

\begin{figure}[t]
\unitlength1cm
\begin{picture}(11,7)
\centerline{\epsfig{file=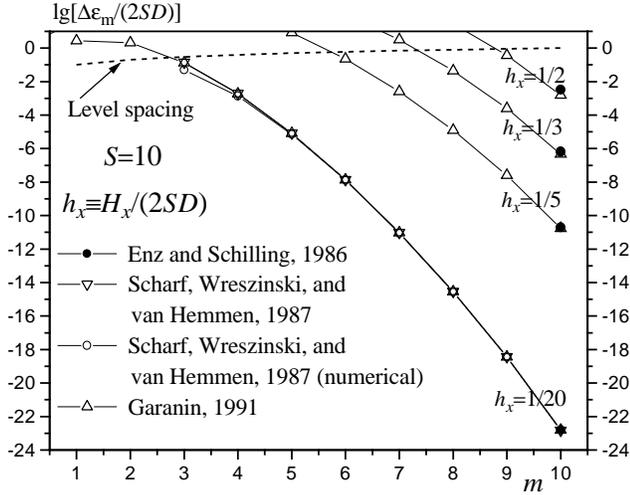,angle=-90,width=12cm}}
\end{picture}
\caption{ \label{t_spl} 
Tunneling splittings $\Delta \varepsilon_{mm'}$ for $H_z=0$ 
and different values of the transverse field. The results of 
Scharf, Wreszinski, and van Hemmen \protect\cite{schwrelvh87} 
are indiscernible from the perturbative ones in this scale 
and they are shown only for the one value of the transverse field. 
}
\end{figure}

The tunneling splittings given by Eq. (\ref{split}) are represented 
in Fig. \ref{t_spl} for $H_z=0$ and
different values of the transverse field, in comparison with the results
of other approaches.
One can see that the splittings change by orders of magnitude with 
changing $m$ by 1.
If the splitting of the pair $m,m'$ becomes comparable to the level spacing
in the well, which is of order $2D|m|$, the tunneling becomes strong and
of {\em nonresonant} character; i.e., the barrier for a particle going
into the other well disappears.
For this pair of levels the perturbation theory clearly breaks down,
but for the next lower pair  $m-1,m'+1$ (see Fig. \ref{t_levs})
it already works well.

The sharp boundary between the levels localized in one of the wells and
the delocalized ones, which was observed above, is also characteristic for the classical theory where there is a similar separation between
the localized and escape orbits at some energy.
Accordingly, as was shown by Friedman, \cite{fridiss} the transverse-field dependence of the classical barrier height,
\begin{equation}\label{classbarrlow}
U(h_x) = DS^2(1-h_x)^2, \qquad h_x \equiv \frac{ H_x }{ 2SD } ,
\end{equation}
can be reproduced for small $h_x$ with the help of the 
perturbative formula (\ref{split}).
Indeed, in the quasiclassical limit $1 \ll |m| \ll S$ Eq. (\ref{split})
for $H_z=0$ can be simplified to
\begin{equation}\label{splitas}
\Delta \varepsilon_{mm'} \cong \frac{ 2D|m| }{ \pi }
\left( \frac{ H_x S e^2 }{ 8Dm^2 } \right)^{2|m|}
\end{equation}
and compared to the level spacing $D|m|$ to obtain the value  
$m=m_b$ at which the barrier is effectively cut by the tunneling.
For $|m|\gg 1$, the value of $m_b$ can be found with a good accuracy 
by equating the 
fraction in brackets in Eq. (\ref{splitas}) to unity.
The result has the form
\begin{equation}\label{mb}
m_b^2\cong 2S^2h_x \frac{e^2}{8} ,
\end{equation}
which leads to the effective barrier height
$U \cong DS^2 - Dm_b^2 \cong DS^2 [1-2h_x(e^2/8)]$.
This is in accordance with Eq. (\ref{classbarrlow}) for $h_x\ll 1$, except
for the factor $e^2/8\approx 0.92$.
The nontrivial feature of this derivation is that the resulting classical
barrier lowering is of first order in $h_x$, although the 
corrections to the energy levels arise only in the second order
of the perturbation theory.
The latter have the form 
\begin{equation}\label{pertenergies}
\varepsilon_m^{(2)} = -Dm^2 \left\{ 1 +
\frac{ 2S^2 [S(S+1)+m^2] }{ m^2 [4m^2-1] } h_x^2 \right\}.
\end{equation}
It can be checked that for $m=m_b$ and $1 \ll m_b \ll S$ the correction term in the curly brackets makes up the universal number $8/e^4\approx 0.15$.
This means that near the renormalized  top of the barrier , $m=m_b$, the perturbation
theory relies on a small {\em numerical} parameter rather than on $h_x$. 
The artifact $e^2/8$ in Eq. (\ref{mb}) is the consequence of 
dropping the effect of the level mixing {\em inside} the wells 
described to lowest order by Eq. (\ref{pertenergies}). 
It should be noted that $e^2/8$ also appears in the WKB results
\cite{lvhsuto86epl,lvhsuto86,schwrelvh87} for the tunneling level splitting
in the case of small transverse field, and it 
can be attributed to the inaccuracy of the WKB method near the top of the 
barrier. 
In this article, we will neglect these effects and  study the tunneling
transitions between the wells perturbatively in the basis of 
the eigenstates of the operator $S_z$.
It should be noted in addition that, as was checked by Chudnovsky and
Friedman, \cite{chufri} the level matching condition  (\ref{hzres}) 
remains uneffected by the transverse field at least up to fourth
order in $h_x$.

Now let us consider the question how the level splitting changes from one
level pair to another in more detail.
For the pairs of resonant levels shown in Fig. \ref{t_levs}, with the use of
the basic formula (\ref{split}) in the unbiased case, one comes to the
result 
\begin{equation}\label{relspl}
\frac{\Delta \varepsilon_{m+1,m'-1} }{\Delta \varepsilon_{mm'} } =
e^4 \left( \frac{ m }{ m_b } \right)^4
\frac{ 
\left( 1 - \frac{1}{2|m|} \right) 
\left( 1 - \frac{1}{|m|} \right) 
}
{ 
\left( 1 + \frac{ |m| }{ S } \right)
\left( 1 - \frac{ |m| }{ S } + \frac 1S \right) 
} ,
\end{equation}
where $m_b$ is given by Eq. (\ref{mb}).
One striking implication of this formula is that the splitting ratio is
large everywhere in the wells: Even near the top of the renormalized  barrier, $m\sim m_b$, the tunneling splitting changes by a large factor 
$e^4\approx 55$, moving one step up the barrier.
This universal behavior, independent of the spin value $S$ for $S\gg 1$,
shows that even in the quasiclassical limit the tunneling splitting cannot be treated as a smooth function of the energy.
The determination of the level at which the barrier disappears is, therefore,
quite precise.
Another consequence of Eq. (\ref{relspl}) is that resonant tunneling
is to the same extent inherent in models of large spin $S$ as in
those of moderate spin.

\section{Spin-bath interactions and the density matrix equation}
\label{dme}

The thermally activated escape of the Mn$_{12}$Ac spin over the 
potential barrier $DS^2\simeq 70$ K is accompanied by the transitions
between the energy levels with the energy differences ranging from
$D(2S-1)\simeq 13$ K near the bottoms of the potential wells to
$D\simeq 0.7$ K near the top of the barrier.
Such a process requires an energy exchange between ${\bf S}$ and 
other degrees of freedom of the whole system.

The dipole-dipole interactions between different magnetic clusters
contribute to the macroscopic magnetic induction 
${\bf B}={\bf H} +4\pi {\bf M}$ which is actually
``felt'' by the spins and which should replace the external field 
{\bf H} in all the formulas for spin tunneling and 
thermal activation.
As was shown in dynamic hysteresis experiments, 
\cite{frisartejzio96,heretal97}
this internal field correction is quite essential for a careful 
analysis of the experimental data.    
The {\em fluctuating} part of the  dipole-dipole interactions
which could cause the spin relaxation has been shown to be inefficient
by diluting the sample. \cite{ses95} 
Indeed, this interaction is of the order of the 
dipole-dipole energy of two neighboring clusters, 
$E_d=(g\mu_B S)^2/v_0$, where $v_0$ is volume of the unit cell.
Using $g=1.9$ and $v_0 = (17.3 {\rm \AA})^2\times 12.4 {\rm \AA} $, 
\cite{lis80} one obtains $E_d\simeq 0.06$ K [in accordance with the 
measured value of the Curie constant $\Theta_C=-0.05$ K
(Ref. \onlinecite{novses95})] which is much smaller
than the distances between the energy levels.
There is also a more subtle argument, \cite{vilharsesret94} based upon
energy conservation and the
nonequidistant character of the spin energy levels,  
which rules out the contribution of dipole-dipole
interactions to the relaxation in the temperature range $T\ll U$.

The nuclear subsystem also cannot supply energies which would be large
enough for the relaxation over the $70$ K barrier.
Nevertheless, nuclear spins produce a hyperfine field on the effective 
electronic spin, which can give rise to tunneling.
This mechanism will be considered in detail in Sec. \ref{numres}.
Here we will describe tunneling as caused by the externally applied
transverse field $H_x$.

The remaining two types of the interaction of a Mn$_{12}$Ac spin with the
environment are those with phonons and photons.
Unlike the interactions reviewed above, the phonon and photon subsystems
play the role of a thermal bath, rendering the spin subsystem a definite
externally controlled temperature.
It can be immediately seen that in the presense of phonons the photon
processes can be safely neglected, since the light velocity $c$ is much
greater than the sound velocity $v$ and, as a result, the photon density of
states is smaller than the phonon one.
At low temperatures the leading processes are the emission and absorption
of phonons, accompanied by the hopping of spin between 
energy levels.
At higher temperatures Raman scattering processes can become dominant.
The energies of phonons in Mn$_{12}$Ac are large enough for the exchange
with the spin subsystem: As follows from specific heat measurements,
\cite{novses95} the Debye temperature $\theta_D$ corresponding to the phonon 
energy at the edge of the Brillouin zone is about 36 K.

Spin-phonon interactions in materials with a strong crystal-field
anisotropy are mainly due to the modulation of the crystal field
by phonons.
This mechanism was extensively studied in past years. \cite{abrble70}
The possible spin-phonon coupling terms for substances of different 
symmetries are listed in Ref. \onlinecite{kol66}.
For Mn$_{12}$Ac and other molecular magnets, the spin-phonon interactions, 
as well as the (presumably complicated) phonon modes themselves, 
have not yet been investigated.
Moreover, an attempt to describe the interaction with phonons rigorously would
lead to a serious complication of the formalism without bringing any new
qualitative results.
We will resort to various simplifications, assuming, in particular,
that the phonon spectra of molecular magnets contain, as for an isotropic
elastic body, one longitudinal and two transverse modes.
Similar simplifications were also made in 
Ref. \onlinecite{vilharsesret94}, where
the pure thermal activation escape rate in Mn$_{12}$Ac was studied.

The lowest-order spin-phonon interactions allowed by the time-reversal 
symmetry are linear in phonon operators and bilinear in the spin operator
components, containing various combinations $S^\alpha S^\beta$, where 
$\alpha,\beta=\pm,z$.
The simplest of these interactions is due to the rotation of the anisotropy
axis by transverse phonons. \cite{harpolvil96} 
We will use this mechanism for the illustration of our method since it does
not employ any unknown characteristics of the crystal-field 
{\em distortions} accompanying other types of lattice vibrations.

For the arbitrarily oriented anisotropy axis {\bf n}, the anisotropy
part of the spin Hamiltonian (\ref{spinham}) can be written as 
${\cal H}_A = -D({\bf nS})^2$.
Transverse phonons change the vector {\bf n} by 
$\delta{\bf n}=[\delta\bbox{\phi} \times {\bf n}]$,
where $\delta\bbox{\phi} = (1/2) \nabla\times {\bf u}$
is the local rotation of the lattice and  {\bf u} is the lattice 
displacement.
The first order term on $\delta{\bf n}$ in ${\cal H}_A$ gives 
the spin-phonon Hamiltonian which in coordinate form reads
\begin{equation}\label{spham}
{\cal H}_{\rm sp} = D\{S_z, S_x\} \omega_{zx} + D\{S_z, S_y\} \omega_{zy} ,
\end{equation}
where
\begin{equation}\label{omrot}
\omega_{\alpha\beta} \equiv \frac 1 2 \left( 
\frac{\partial u_\alpha }{\partial r_\beta } -
\frac{\partial u_\beta }{\partial r_\alpha } \right) ,
\end{equation}
and $\{S_\alpha, S_\beta\}$ is the anticommutator.
In terms of phonon operators $a_{k\lambda}$ and $a_{k\lambda}^\dagger$,
\begin{equation}\label{phonopers}
{\bf u} = \frac{ i  }{ (2MN)^{1/2} }
\sum_{{\bf k}\lambda} 
\frac{ {\bf e}_{{\bf k}\lambda} e^{i {\bf kr}} }
{ (\omega_{k\lambda})^{1/2} } (a_{k\lambda}- a_{k\lambda}^\dagger ) , 
\end{equation}
where $M$ is the unit cell mass, $N$ is the number of cells in the lattice,
${\bf e}_{{\bf k}\lambda}$ is the phonon polarization vector, 
$\lambda=t,t,l$ is the polarization, and $\omega_{k\lambda}=v_\lambda k$
is the phonon frequency.
Performing differentiation in Eq. (\ref{omrot}),
one can transform Eq. (\ref{spham}) to
\begin{equation}\label{sphamaa}
{\cal H}_{\rm sp} = - \frac{ 1 }{ N^{1/2} } \sum_{{\bf k}\lambda} 
V_k \{S_z, (\bbox{\eta}_{{\bf k}\lambda} {\bf S}) \}
(a_{k\lambda}- a_{k\lambda}^\dagger ) .
\end{equation}
Here the spin-phonon amplitide $V_k$ is given by
\begin{equation}\label{vk}
V_k = \frac{ D }{ 2^{3/2} } \left( 
\frac{ \omega_{k \rm t} }{ \Omega_t } \right)^{1/2},
\qquad 
\Omega_t \equiv M v_t^2 ,
\end{equation}
and the vector $\bbox{\eta}_{{\bf k}\lambda}$ is determined by
\begin{equation}\label{etvect}
\eta_{{\bf k}\lambda}^z = 0,
\qquad
\eta_{{\bf k}\lambda}^{x,y} =  
e_{{\bf k}\lambda}^z n_{\bf k}^{x,y} - 
e_{{\bf k}\lambda}^{x,y} n_{\bf k}^z ,
\end{equation}
where ${\bf n}_{\bf k} \equiv {\bf k}/k$. 
On can see that the coupling to longitudinal phonons in Eq.
(\ref{sphamaa}) vanishes, as it should be, since 
${\bf e}_{{\bf k}\rm l} = {\bf n}_{\bf k}$.

The evolution of a spin system coupled to an equilibrium heat bath can be
described by the density matrix equation.
The diagonal elements of the density matrix, $\rho_{mm}\equiv N_m$, describe
the population of the energy levels.
In the absense of interactions noncommuting with $S_z$ in the spin Hamiltonian
$\cal H$, the DME reduces to the closed system of kinetic balance equations,
or master equations, for the populations $N_m$ in the basis of the eigenstates of the operator $S_z$.
The latter was applied to describe the thermoactivation
process in uniaxial spin systems, as Mn$_{12}$Ac, in Refs. \onlinecite{vilharsesret94} and \onlinecite{gar97pre}.
If a transverse field or another level mixing perturbation is applied to
the system, the nondiagonal elements of the DME appear,
whose slow dynamics describes the tunneling process.      
The major advantage of the DME is that it provides a natural account of 
resonant tunneling in systems of moderate spin, which is lost in
quasiclassical approaches for truly macroscopic systems.

A common routine for obtaining a system of kinetic balance equations
is to calculate the transition probabilities according the Fermi
golden rule and then insert them into the equations that are themselves
postulated but not derived.
Such an approach is methodically insufficient since the transition
probabilities are obtained with the help of the time dependent
perturbation theory where the probability of finding the system in 
states differing from the initial fully occupied state are used as a small
parameter.
In other words, this method describes only the initial stage of the
relaxation process for a special type of initial conditions.
Although it incidently leads to the correct master equation, the same is
not true for the general DME.
Indeed, spin-phonon couplings of the type 
$\sum_k \Psi_k S_z^2 a_k a_k^\dagger$, 
corresponding to the elastic scattering of phonons,
do not result in transitions between the energy levels and do not
contribute to the coefficients of the master equation.
On the other hand, such terms modulate the energy levels and contribute to
the linewidths, which manifest themselves in the dynamics of the nondiagonal
elements of the density matrix.

A rigorous method of the derivation of the density matrix equation valid for 
all times employs the projection operator technique. 
\cite{nak58,zwa60,zwa61,gra82}
For spin systems, the details of calculations are described in Ref. \onlinecite{romorsopp89}.
The resulting DME can be found in 
Ref. \onlinecite{gar91llb}, where the model without single-site
anisotropy, accounting for both one-phonon and Raman scattering processes,
was used to derive the Landau-Lifshitz-Bloch equation for ferromagnets.
This DME is written in terms of the Hubbard operators 
$X^{mn}\equiv |m\rangle \langle n|$ forming the 
complete basis for the spin subsystem.
In the Heisenberg representation the operators $X^{mn}$ are related to the
spin density matrix: $\rho_{mn}=\langle X^{mn}(t) \rangle$.
For the present model described by Eqs. (\ref{spinham}) and (\ref{sphamaa}),
the resulting DME reads
\begin{eqnarray}\label{dmeeq}
&&
\dot X^{mn} =  i  \omega_{mn} X^{mn} -
\frac{ i  }{ 2 } H_x 
( l_{m,m+1} X^{m+1,n} + l_{m,m-1} X^{m-1,n} 
\nonumber\\
&&
\qquad
- l_{n,n+1} X^{m,n+1} - l_{n,n-1} X^{m,n-1} )
+ R_{mn},
\end{eqnarray}
where $\omega_{mn}\equiv \varepsilon_m - \varepsilon_n$ are the frequencies
associated with the transition $n\to m$, the unperturbed energy levels 
$\varepsilon_m$ are given by $\varepsilon_m = -Dm^2 - H_zm$, the factors
$g\mu_B$ and $\hbar$ are dropped for convenience, the matrix elements
$l_{m,m\pm 1}$ are given by Eq. (\ref{lm}), and $R_{mn}$ is the relaxation term.
The latter has the non-Markovian form
\begin{equation}\label{rmn}
R_{mn} = - \int_{t_0}^t dt' \frac 1N \sum_{{\bf k}\lambda} V_k^2
\{ A f_k(t'-t) - B f_k(t-t') \} ,
\end{equation}
where
\begin{eqnarray}\label{rmnab}
&&
A = Q_S(t') [Q_S(t), X^{mn}(t) ],
\nonumber\\
&&
B = [ Q_S(t), X^{mn}(t) ] Q_S(t'),
\end{eqnarray}
the spin operator combination 
$Q_S \equiv \{S_z, (\bbox{\eta}_{{\bf k}\lambda} {\bf S}) \}$
comes from the spin-phonon Hamiltonian (\ref{sphamaa}),
the function $f_k(\tau)$ characterizing the bath in the present case of 
the one-phonon processes is given by
\begin{equation}\label{rmnfk}
f_k(\tau) = n_k e^{ i  \omega_k\tau} 
+  (n_k+1) e^{ -i  \omega_k\tau},
\end{equation}
$n_k \equiv (e^{\omega_k/T} - 1)^{-1}$ are the boson occupation
numbers, and  $\omega_k\equiv \omega_{kt}$ are the frequencies of
transverse phonons.

In Eq. (\ref{rmnab}) the spin operators should be expanded over the $X^{mn}$
basis as follows:
\begin{eqnarray}\label{sxmn}
&&
S_+ = \sum_{m=-S}^{S-1} l_{m,m+1} X^{m+1,m}, 
\qquad 
S_z = \sum_{m=-S}^S m X^{mm} , 
\nonumber\\
&&
S_- = \sum_{m=-S}^{S-1} l_{m,m+1} X^{m,m+1},
\qquad
S_\pm \equiv S_x\pm i  S_y .  
\end{eqnarray}
For the one-phonon processes, 
the integral over $t'$ in Eq. (\ref{rmn}) converges on the scale of
$1/\omega_{mn}$ which is
much shorter than the relaxation time of the spin system.
Hence, the lower limit of this integral can be extended to $t_0=-\infty$
and the $t'$ dependences of the operators $X^{mn}$ in the relaxation term 
can be considered as governed solely by the conservative part of the 
DME (\ref{dmeeq}).
Finding these time dependences is a matter of numerical work, if the
transverse field $H_x$ is not small.
Here serious complications arise, since the evolution of each operator
$X^{mn}$ is a linear combination of all possible types of spin motion.
This means simply that the unperturbed basis we have chosen 
is not suitable in situations with strong level mixing.
However, in the case of small $H_x$ one can neglect these effects and 
use the unperturbed time dependences
\begin{equation}\label{xmnt}
X^{mn}(t') = e^{i  \omega_{mn}(t'-t) } X^{mn}(t) .
\end{equation}

Now one can calculate combinations $A$ and $B$ in Eq. (\ref{rmnab}) with
the use of the representations (\ref{sxmn}) and the equal-time relation
$X^{mk} X^{ln} = X^{mn} \delta_{kl}$    
which replaces the commutation relations for the spin components.
The sum over the phonon polarizations $\lambda$ in Eq. (\ref{rmn}) can be
done using Eq. (\ref{etvect}) and the property of the polarization vectors
$\sum_\lambda e_\lambda^\alpha e_\lambda^\beta = \delta_{\alpha\beta}$.
Neglecting the imaginary part of the relaxation term $R_{mn}$, corresponding
to the renormalization of the spin energy levels due to the coupling to
the bath, one arrives at the final form of $R_{mn}$:
\begin{eqnarray}\label{rmnfin}
&&
R_{mn} = \frac 12 \bar l_{m,m+1} \bar l_{n,n+1} 
[W_{m,m+1} + W_{n,n+1}] X^{m+1,n+1}
\nonumber\\
&&
\qquad
{} - \frac 12 [\bar l_{m,m+1}^2 W_{m+1,m} + 
\bar l_{n,n+1}^2 W_{n+1,n}] X^{mn}
\nonumber\\
&&
\qquad
{} + \frac 12 \bar l_{m,m-1} \bar l_{n,n-1} [W_{m,m-1} + W_{n,n-1}]
X^{m-1,n-1}
\nonumber\\
&&
\qquad
{} - \frac 12 [\bar l_{m,m-1}^2 W_{m-1,m} + \bar l_{n,n-1}^2 W_{n-1,n}]
 X^{mn}.
\end{eqnarray}
Here $\bar l_{m,m\pm 1} \equiv l_{m,m\pm 1} (2m\pm 1)$ with the factor
$2m\pm 1$ coming from the 
operator $S_z$ in Eq. (\ref{sphamaa}), and the universal rate constant
$W_{mn}=W(\omega_{mn})$ of the one-phonon processes is given by
\begin{eqnarray}\label{w1}
&&
W(\omega) = \frac 23 v_0 \!\!\int\!\! \frac{ d{\bf k} }{(2\pi)^3 } V_k^2 
\big\{ (n_k+1) \pi\delta(\omega_k+\omega) \nonumber \\
&&\qquad\qquad\qquad\qquad
{} + n_k \pi\delta(\omega_k-\omega) \big\} ,
\end{eqnarray}
where $v_0$ is the unit cell volume and the overall factor 2/3 says
that only two transverse modes of the total three phonon
modes are active in the relaxation mechanism under consideration.
One can check that the rate constant satisfies the detailed 
balance condition $W(\omega)=W(-\omega)\exp(-\omega/T)$.
At low temperatures phonons die out and $W(\omega)$ with $\omega>0$, which 
corresponds to the absorption of a phonon, becomes exponentially small.
The result for $W(\omega)$ with $\omega<0$ (the emission of a phonon) 
calculated with the help of Eqs. (\ref{w1}) and (\ref{vk}) reads 
\begin{equation}\label{w1res}
\renewcommand{\arraystretch}{1.2}
W = \frac{D^2|\omega|^3}{24\pi\Theta^4}(n_{|\omega|}+1)
\propto 
\left\{
\begin{array}{ll}
\omega^2 T, & |\omega|\ll T
\\
|\omega|^3, & T\ll |\omega|
\end{array}
\right.
\end{equation}
(cf. Ref. \onlinecite{orb61}).
Here we have used $\theta_D^3 = (\hbar v_t )^3/v_0$ for the Debye temperature 
$\theta_D\sim\hbar\omega_{k_{\rm max}}$.
The constant $\Theta$ is defined as   
$\Theta^4\equiv\Omega_t\theta_D^3 = \hbar^3\rho^2 v_t^5$, 
where $\rho$ is the density and $\Omega_t$ is given by Eq. (\ref{vk}).

Note that Eq. (\ref{dmeeq}) with $R_{mn}$ given by Eq. (\ref{rmnfin})
is still an operator equation, and the equation of motion for the density
matrix elements, $\rho_{mn} \equiv \langle X^{mn} \rangle$, should be obtained
by taking its quantum-statistical average over the initial state of the spin.
This is, however, a trivial task, since the equation for $X^{mn}$ is linear.

In the case $H_x=0$ the density matrix equation (\ref{dmeeq}) and
(\ref{rmnfin}) reduces to a 
system of kinetic balance equations for the diagonal elements 
$N_m\equiv X^{mm}$, the equilibrium solution of which is given by
\begin{equation}\label{n0}
N_m^{(0)} = \frac{1}{Z} e^{-\varepsilon_m/T},
\qquad Z =\!\! \sum_{m=-S}^S \!\!e^{-\varepsilon_m/T} .
\end{equation}
The thermoactivation relaxation rate $\Gamma$ in the model with $H_x=0$
was studied in Ref. \onlinecite{vilharsesret94} and recently in 
Ref. \onlinecite{gar97pre}.
In the latter work Raman scattering processes have also been taken into 
account, and the spin relaxation rate was calculated for arbitrary ratios
$U/T$ in terms of the integral relaxation time $\tau_{\rm int}$.
It was shown that in systems with larger spin values, even in the Arrhenius
regime $U/T\gg 1$, there are several limiting cases for the prefactor
$\Gamma_0$ in the expression $\Gamma=\Gamma_0\exp(-U/T)$ 
as a result of the interplay between the one-phonon and Raman scattering
processes.
Here we concentrate on the low-temperature region, and thus only one-phonon
processes will be considered.

\section{Slow dynamics of the density matrix: coherence and tunneling 
between resonant levels}
\label{slowdyn}

The possible frequencies, with which the density matrix elements $X^{mn}$
evolve in time according to the DME (\ref{dmeeq}), range from 
$\omega_{0S}=DS^2$ (for $H_z=0$) to very small ones corresponding to
overbarrier relaxation and tunneling.
In the low-temperature range $T\ll U$, these fast motions decay with the 
rate corresponding to the relaxation inside one well, which is much larger 
than the thermoactivation escape rate or the tunneling rates.
In the long-time or low-frequency dynamics, the variables $X^{mn}$
corresponding to the large $\omega_{mn}$ play the role of ``slave''
degrees of freedom, adjusting themselves to the evolution of the slow
variables, and hence they can be adiabatically eliminated.

The slow variables of our problem are the diagonal matrix elements 
$N_m=X^{mm}$, as well as the antidiagonal elements $X^{mm'}$ whose
transition frequency 
$\omega_{mm'}\equiv \varepsilon_m - \varepsilon_{m'}$ 
is the detuning of the resonant levels
$m$ and $m'$
\begin{equation}\label{detun}
\omega_{mm'} = (H-H_k)(m'-m)
\end{equation}
[cf. Eqs. (\ref{hzres}) and (\ref{epsres})].
The equations of motion for these slow variables can be obtained in the following way.
In Eq. (\ref{dmeeq}) for $X^{mm}$, the terms containing $X^{m+1,m}$
and $X^{m,m+1}$, which are generated minimally by nonzero $X^{m'm}$ and $X^{mm'}$, correspondingly, are responsible for tunneling
in the lowest approximation.
In the dynamical equations for these elements one can neglect the terms
$\dot X^{m+1,m}$ and $\dot X^{m,m+1}$, as well as the relaxation terms,
since the frequencies $\omega_{m+1,m}$ and $\omega_{m,m+1}$ are large on
the scale of relaxational and tunneling processes.
Then, in the case of $X^{m+1,m}$, this element can be expressed with the 
help of its dynamical equation through $X^{m+2,m}$ as   
\begin{equation}\label{xiter}
X^{m+1,m} = \frac{ \frac 12 H_x l_{m,m+1} }{\omega_{m+1,m} } X^{m+2,m} . 
\end{equation}
In the right part of this equation the terms containing $X^{mm}$,
$X^{m+1,m+1}$, and $X^{m+1,m-1}$ have been dropped because retaining them
would be against our strategy of going across the barrier along the
shortest path to $X^{m'm}$.
For the same reason we have also dropped the terms $X^{m-1,m}$ and
$X^{m,m-1}$ in the equation for $X^{mm}$.
Retaining all these terms would imply taking into account the level mixing
inside the wells, which we neglect for small transverse
fields.         
Now, Eq. (\ref{xiter}) can be iterated until $X^{m+1,m}$ is expressed
through $X^{m'm}$, and similar can be performed on $X^{m,m+1}$.
Substituting their expressions into the equation for $X^{mm}$, one arrives
at the slow equation
\begin{equation}\label{xmmslow}
\dot X^{mm} = \frac{ i  }{ 2 } \Omega_{mm'} ( X^{mm'} - X^{m'm} ) 
+ R_{mm} ,  
\end{equation}
where $\Omega_{mm'}$ is the tunneling frequency coinciding with the 
tunneling level splitting $\Delta\varepsilon_{mm'}$ of Eq. (\ref{split}).
One can see now that the algorithm used here for the adiabatic elimination 
of the fast degrees of freedom in the density matrix equation is the dynamic
counterpart of the perturbative approach leading to the chain formula
(\ref{chain}).   
The antidiagonal matrix elements $X^{mm'}$ and $X^{m'm}$ are generated,
in turn, by the diagonal elements $X^{mm}$ and $X^{m'm'}$, and the dynamical
equations for them can be obtained in a similar way.
The result for $X^{mm'}$ reads
\begin{equation}\label{xmm'slow}
\dot X^{mm'} = i  \omega_{mm'} X^{mm'} 
- \frac{ i  }{ 2 } \Omega_{mm'} ( X^{m'm'} - X^{mm} ) 
+ R_{mm'} .   
\end{equation}
For the matrix elements $X^{m'm'}$ and $X^{m'm}$ one obtains equations
similar to Eqs. (\ref{xmmslow}) and (\ref{xmm'slow}).

To formulate the resulting system of slow equations in a more convenient
form, we introduce 
\begin{eqnarray}\label{tlsvar}
&&
Z_{mm'} \equiv N_{m'} - N_m , 
 \nonumber\\
&&
Y_{mm'} \equiv i  ( X^{mm'} - X^{m'm} ),
 \nonumber\\
&&
X_{mm'} \equiv   X^{mm'} + X^{m'm} .
\end{eqnarray}
These variables satisfy the system of equations
\begin{eqnarray}\label{ncoheq}
&&
\dot N_m = \frac 12 \Omega_{mm'} Y_{mm'} + R_{mm} ,   
 \nonumber\\
&&
\dot N_{m'} = -\frac 12 \Omega_{mm'} Y_{mm'} + R_{m'm'}
\end{eqnarray}
[cf. Eq. (\ref{xmmslow})]   
and
\begin{eqnarray}\label{tlseq}
&&
\dot Z_{mm'} = - \Omega_{mm'} Y_{mm'} + R_{m'm'} -  R_{mm}, 
 \nonumber\\
&&
\dot Y_{mm'} = \Omega_{mm'} Z_{mm'} - \omega_{mm'} X_{mm'}
- \Gamma_{mm'} Y_{mm'},
 \nonumber\\
&&
\dot X_{mm'} = \omega_{mm'} Y_{mm'} 
- \Gamma_{mm'} X_{mm'},
\end{eqnarray}
where the equation first of Eqs. (\ref{tlseq}) is a consequence
of Eqs. (\ref{ncoheq}).
The conservative part of Eqs. (\ref{tlseq}) describes the precession of the 
pseudospin $\bbox{\sigma}_{mm'} \equiv \{X_{mm'}, Y_{mm'}, Z_{mm'}\}$ in
the pseudofield ${\bf H}_{mm'} \equiv \{\Omega_{mm'}, 0, \omega_{mm'}\}$.
In the absense of dissipation, in resonance ($\omega_{mm'}=0$), the
pseudospin rotates in the $y,z$ plane, and the difference of the level
populations $Z_{mm'}$ oscillates with time.
Note, however, that the $Y$ and $X$ components of the pseudospin
have nothing to do with the actual spin components $S_y$ and $S_x$ which
remain zero, see Eq. (\ref{sxmn}).
The only exclusion is the resonance between the two neighboring levels 
$m$ and $m+1$ near
the  top of the barrier , which is realized, e.g., for $S$ odd and $H_z=0$.
In this case, which is actually no longer the tunneling case since 
$\Omega_{m,m+1} \propto H_x$ is not suppressed by the anisotropy,
the rotation of the pseudospin couples to the rotation of the real spin.

Since the tunneling frequency  $\Omega_{mm'}$ is typically very small, the 
correspondingly small detuning $\omega_{mm'} \geq \Omega_{mm'}$ 
[see Eq. (\ref{detun})] is sufficient to suppress the resonance. 
On the other hand, a small ac field $H_z(t)$ with a frequency about
$\Omega_{mm'}$ giving rise to the corresponding $z$ component of the 
pseudofield $\omega_{mm'}(t)$ [see Eq. (\ref{detun})] can excite the 
tunneling resonance.
The latter, however, can only happen under two rather severe conditions:
\begin{equation}\label{rescond}
H_z(t) \ll \frac{ \hbar \Omega_{mm'} }{ (m'-m) },
\qquad T\ll \hbar \Omega_{mm'} .
\end{equation}
The former is the condition of the linear resonance, whereas the latter 
requires that the pseudospin have a strong preference along the $x$ axis,
in other words, that only the lower of the tunneling-splitted states 
(the even one) is thermally populated.
The temperatures required by the second condition are so small that only
the resonance between the ground-state levels $m=\pm S$ can be discussed.

The small value of the pseudofield $\Omega_{mm'}$ in resonant tunneling
equations (\ref{tlseq}) suggests an important role of the relaxation terms.
The diagonal relaxation term $R_{mm}$ following from Eq. (\ref{rmnfin})
has the form
\begin{eqnarray}\label{rmm}
&&
R_{mm}= \bar l_{m,m+1}^2 ( W_{m,m+1} N_{m+1} - W_{m+1,m} N_m )
\nonumber \\
&&\qquad
{} + \bar l_{m,m-1}^2 ( W_{m,m-1} N_{m-1} - W_{m-1,m} N_m ) ,
\end{eqnarray}
describing the exchange of particles with the levels $m\pm 1$.
For the antidiagonal matrix elements, the relaxation term $R_{mm'}$
in Eq. (\ref{rmnfin}) contains $X^{mm'}$ itself, as well as the matrix
elements $X^{m\pm 1, m'\pm 1}$.
These matrix elements do not belong, however, to the antidiagonal ones
(see Fig. \ref{t_levs}); they are small slave variables that have
been eliminated above.
Dropping them leads to
\begin{eqnarray}\label{gammm'}
&&
\Gamma_{mm'} = \Gamma_m + \Gamma_{m'},
\nonumber\\
&&
\Gamma_m = \frac 12 ( \bar l_{m,m+1}^2 W_{m+1,m} + \bar l_{m,m-1}^2 W_{m-1,m} ) .
\end{eqnarray}
Here the terms $\Gamma_m$ and the analogous $\Gamma_{m'}$ are the linewidths
of the levels $m$ and $m'$ arising from the transitions to the levels
$m\pm 1$ and $m'\pm 1$ with the absorption or emission of an energy quantum.

At temperatures $T \ll \omega_{\rm A} = (2S-1)D$, which is about 
13 K for Mn$_{12}$Ac, most of the particles are in the ground states $m=\pm S$.
The linewidths of these states are much smaller than that of excited ones
since in Eq. (\ref{gammm'}) the emission term is absent and the absorption
term is small as $\exp(-\omega_{\rm A}/T)$.
Further lowering of the temperature leads to the suppression of the thermoactivation
relaxation mechanism and, simultaneously, to the vanishing of dissipation in
the ground state.
Thus, the spin of the magnetic cluster behaves like an undamped two-level
system (TLS).
It is, however, well known (see, e.g., Ref. \onlinecite{legetal87}) that the 
coupling of the TLS to the bath strongly changes its dynamics,
and one can ask where this coupling was lost in our calculations.
The answer is that treating the non-Markovian relaxation term (\ref{rmn}) 
we have used the simplest unperturbed $t'$ dependences (\ref{xmnt}) for the
spin operators of Eqs. (\ref{rmnab}) and (\ref{sxmn}),
which do not describe the tunneling motion.
This tunneling motion couples, however, to a very small number of 
extremely-long-wavelength phonons, and their contribution to the relaxation terms
is smaller by a factor of order  
$(\Omega_{-S,S}/\omega_{\rm A})^3 \exp(\omega_{\rm A}/T)$
[see Eq. (\ref{w1res})]
than that of the regular phonon processes.
Thus, the coupling of the tunneling mode to the bath becomes important 
only at very low temperatures.
In this range serious complications arise 
(see, e.g., Ref. \onlinecite{legetal87})
since the pseudospin part of the effective TLS Hamiltonian,  
${\cal H}_{\rm TLS} = -\bbox{\sigma\Omega}$, is no longer large in
comparison to the coupling to the bath and the perturbation theory breaks
down.

The equation of motion for the pseudospin, Eq. (\ref{tlseq}), is not closed 
because the relaxation term in the first line couples it to other levels.
If we neglect this coupling for a moment, then the eigenvalues $\lambda$ of 
Eq. (\ref{tlseq}) determined as $X,Y,Z \propto e^{-\lambda t}$
are given by the roots of the cubic equation
$(\lambda-\Gamma)^2\lambda + \Omega^2(\lambda-\Gamma) + \omega^2\lambda=0$,
where we have dropped the index $mm'$.
This equation can be solved only in limiting cases.
In particular, in resonance ($\omega=0$) the last equation of 
Eqs. (\ref{tlseq})
decouples from the first two ones, which describe now a damped harmonic
oscillator with $\lambda_{1,2}=(1/2)(\Gamma \pm \sqrt{\Gamma^2-4\Omega^2})$.
One can see that the tunneling oscillations of the particle between the two
levels become overdamped for $\Gamma>2\Omega$.
In the small damping case, the solution of Eq. (\ref{tlseq}) with the
initial condition $Z(0)=1$ has an interesting two-scale-relaxation form
\begin{eqnarray}\label{tlssol}
&&
Z(t) = \frac{ \Omega^2 }{ \Omega^2 + \omega^2 }
\exp\left(-\frac{ \Omega^2/2 + \omega^2 }
{\Omega^2 + \omega^2} \Gamma t \right) \cos(\sqrt{\Omega^2 + \omega^2}t) 
 \nonumber\\
&&
{} + \frac{ \omega^2 }{ \Omega^2 + \omega^2 }
\exp\left(-\frac{ \Omega^2 }{ \Omega^2 + \omega^2} \Gamma t \right) .
\end{eqnarray}
These results should not be overstated for the present model because in the
underdamped case the neglected relaxation terms in the equation for $Z$ can
be of the same order of magnitude as the accounted ones in the equations for
$X$ and $Y$.
In this case the pseudospin concept breaks down and one should use the two
equations (\ref{ncoheq}) instead of the first equation of Eqs. (\ref{tlseq}).
But in the case of strong damping the level populations cannot deviate 
substantially from their equilibrium values because of the slow
tunneling motion, and the different terms in the diagonal relaxation terms
$R_{mm}$ given by Eq. (\ref{rmm}) nearly cancel each other.
Here the concept of the independent pseudospin is justified, and one can see
that its motion is indeed overdamped.
Neglecting the terms $\dot X$ and $\dot Y$ in Eqs. (\ref{tlseq}), one
eliminates $X$ and $Y$ and comes to the simple relaxational equation for
$Z$ with $\lambda=\Omega^2\Gamma/(\omega^2+\Gamma^2)$.

The argument in favor of the pseudospin model is that there can be
other relaxation mechanisms, such as those due to spin-spin interactions,
which contribute only to the linewidths (i.e., to the transverse relaxation 
rate) and not to the transition probabilities (i.e., to the longitudinal
relaxation rate). 
In this typical for the magnetic resonance situation  
the term $R_{m'm'}-R_{mm}$ in the first equation of Eqs. (\ref{tlseq}) can be
neglected on the relatively short scale of the transverse relaxation time.
In our model the dipole-dipole interactions could play such a role, but for
Mn$_{12}$Ac the main effect of such a type comes from nuclear spins
(see Sec. \ref{numres}).

The possibility of the overdamping of the coherent spin oscillations was
pointed out by Garg, \cite{garg95diss} who considered resonant
tunneling with the help of a phenomenological damped Schr\"odinger equation
in the matrix representation in the unperturbed basis. 
Although the qualitative conclusions of Garg are the same as the present 
ones, there are some discrepancies between the two approaches in treating the
relaxation.
In particular, the eigenvalues for the two-level problem satisfy
in Garg's approach a quadratic equation instead of the cubic or quartic
ones in our method.
Garg's solution for the splitted energy levels
$\tilde\varepsilon_{1,2}$ is explicitly given by 
$\tilde\varepsilon_{1,2}=(1/2)[E_1 + E_2 \pm \sqrt{(E_1 - E_2)^2 + \Omega^2}]$,
where $E_i \equiv \varepsilon_i - i  \Gamma_i$ are the 
damped ``unperturbed'' energy levels.
Here the well-known deficiency of the damped Schr\"odinger equation can be
seen: The linewidths of the two levels cancel each other under
the square root which is responsible for the tunneling.
In the symmetric (unbiased) case this cancellation is complete, and the
tunneling resonance cannot be overdamped, in contrast to the results of the
density
matrix formalism where the linewidths are added [see Eq. (\ref{gammm'})].
This problem was avoided by Garg by considering the resonance between the
zero-width ground-state level in one well with an excited one in the other
well in the low-temperature biased case, which allowed him to obtain
plausible results.
In our model in this case one should use Eqs. (\ref{ncoheq}) with 
$R_{mm}=0$ and $R_{m'm'}=-2\Gamma_{m'}N_{m'}$, as well as the second and
the third equations of Eqs. (\ref{tlseq}) with $\Gamma_{mm'}=\Gamma_{m'}$,
which leads to a quartic secular equation for $\lambda$. 
In fact, however, such tunneling resonances are typically overdamped, and
both methods give the same results.
The coherent tunneling oscillations should be looked for between the two
ground-state levels whose damping is very small.
For this situation, as well as for the description of thermally activated
tunneling, the damped Schr\"odinger
equation is inappropriate even as a qualitative tool.

In the Arrhenius regime the rate
of the process is controlled by the climbing of particles up the barrier,
which is small in comparison to $\Gamma_{mm'}$ of Eqs. (\ref{gammm'}).
In this case, again, one can neglect the time derivatives 
$\dot X$ and $\dot Y$ in Eqs. (\ref{tlseq}), which leads to the system of
balance equations
\begin{equation}\label{mastereq}
\dot N_m = \frac{ \Omega_{mm'}^2 }{2} 
\frac{ \Gamma_{mm'} }{ \omega_{mm'}^2 + \Gamma_{mm'}^2 } (N_{m'} - N_m)
+ R_{mm},   
\end{equation}
where the rate coefficient for the transition across the barrier is the same
as in the overdamped case and $R_{mm}$ is given by Eq. (\ref{rmm}).
The form of these equations is quite plausible and resembling of the Fermi golden rule: The tunneling frequency $\Omega$ is the
transition amplitude [cf. Eq. (\ref{split})], whereas 
$\Gamma_{mm'}/( \omega_{mm'}^2 + \Gamma_{mm'}^2)$ plays the role of the 
$\delta$ function selecting the allowed resonant level partners.
In our case of the discrete spectrum, one cannot set the latter to the  
$\delta$ function, which causes a small problem:  If the two levels are
not exactly in resonance, the tunneling term prevents establishing the 
equilibrium Boltzmann distribution (\ref{n0}). 
The corresponding deviations from the equilibrium are, however,
small and they can be neglected, especially as we ignore all the 
effects of the level renormalization due to the transverse field.
More important is that the tunneling term in Eq. (\ref{mastereq}) allows
the establishing of the equilibrium between the two wells by crossing the
barrier, and this process is of resonant character.
One can speculate how the form of this term manifests itself in the 
escape rate $\Gamma$ and what will be the shape of the
corresponding resonances.
These questions will be answered in the next section.

\section{Escape rate in the thermally activated regime}
\label{kram}

As was said at the end of the previous section, in the low-temperature
range $T\ll U$ the rate of thermal activation to the top of the barrier is
much lower than that of the relaxation between the neighboring levels.
In this situation quasiequilibrium is promptly established in each of the
wells, and the subsequent relaxation changes only the collective variables 
--- the numbers of particles in the wells, $N_\pm$.
On this stage the problem can be solved analytically, and the solution 
shows that deviations from quasiequilibrium are localized to the 
narrow region near the top of the barrier.
For the thermal activation of particles described by the Fokker-Plank equation,
this problem was solved in the pioneering work of Kramers. \cite{kra40}
The same method was applied later to classical magnetic particles by Brown.
\cite{bro63} 
For the spin system with a discrete spectrum the generalization was given
in Ref. \onlinecite{vilharsesret94}.
Another method applicable in the whole temperature range, for small
deviations from equilibrium, was suggested
in Refs. \onlinecite{garishpan90} and \onlinecite{gar96pre} for classical 
magnetic particles
and in Ref. \onlinecite{gar97pre} for discrete spin systems.

In our low-temperature case, the time derivatives in Eq. (\ref{mastereq})
can be neglected for all values of $m$ except for those near the bottom of
the wells, practically except for $m=\pm S$.
This is because the thermal activation process is exponentially slow and,
in addition, the level populations away from the bottoms are exponentially
small.
Now let us represent $N_m$ in  Eq. (\ref{mastereq}) as
\begin{equation}\label{umdef}
N_m \equiv  N_m^{(0)} u_m ,
\end{equation}
where $N_m^{(0)}$ is the equilibrium population of the level $m$ given by
Eq. (\ref{n0}) and $u_m$ describes deviations from equilibrium.
In terms of $u_m$ the kinetic equation (\ref{mastereq}) can be with the use
of Eq. (\ref{rmm}) rewritten as
\begin{eqnarray}\label{kirch}
&&
0 = j_{mm'} + j_{m,m+1} + j_{m,m-1},
\nonumber\\
&& 
j_{mn} = \sigma_{mn} (u_n - u_m) , 
\end{eqnarray}
where $j_{mn}$ has the meaning of the particle's current from the $n$th to
the $m$th level, $u_m$ plays the role of a potential, and the conductances
$\sigma_{mn}$ are given by
\begin{eqnarray}\label{sigmn}
&&
\sigma_{mm'} \equiv \frac{ \Omega_{mm'}^2 }{2} 
\frac{ \Gamma_{mm'} }{ \omega_{mm'}^2 + \Gamma_{mm'}^2 } 
\frac{ N_m^{(0)}+ N_{m'}^{(0)} }{ 2 } ,
\nonumber\\
&& 
\sigma_{m,m+1} \equiv \bar l_{m,m+1}^2 W_{m+1,m} N_m^{(0)} ,
\end{eqnarray}
where for the tunneling process we have dropped the small terms violating the
equilibrium Boltzmann distribution and symmetrized the rest.
One can check that $\sigma_{m,m+1} = \sigma_{m+1,m}$ due to the symmetry of 
$\bar l_{m,m+1}$ and the detailed balance condition
$ W_{m+1,m} N_m^{(0)} =  W_{m,m+1} N_{m+1}^{(0)}$.
In the high-barrier limit $T\ll U$ the quantities $\sigma_{m,m+1}$ are 
determined mainly by the Boltzmann factors and they become very small near
the top of the barrier.
On the contrary, for not too low temperatures the tunneling conductances
$\sigma_{mm'}$ are extremely small near the bottom and increase by a giant
factor [see Eq. (\ref{relspl})] with each step to the top of the barrier.
As a result, $\sigma_{mm'}$ is essential only near the  top of the barrier , where it
competes with $\sigma_{m,m+1}$ and shunts the equivalent resistor circuit.

In a broad range of $m$ not close to either the top or the bottom the particle's
currents $j_{m,m-1}$ in both wells are practically constant and equal to
each other; let us denote them $j_{m,m-1}\equiv j_{+-}$, the current from
the left ($-$) to the right (+) well.
Then one can write
\begin{equation}\label{nderiv}
\dot N_+ = j_{+-} , \qquad  \dot N_- = -j_{+-} ,
\end{equation}
for the numbers of particles in both wells.
The potential $u_m$ is also constant in the main part of the wells and  changes near the top of the barrier where $\sigma_{m,m+1}$ are especially
small, in accordance with the concept of quasiequilibrium described above.
Denoting the values of $u$ in the wells as $u_+$ and $u_-$, 
one can relate the difference $u_+ - u_-$  to the particle's current 
$j_{+-}$ by the linear relation
\begin{equation}\label{ohm}
j_{+-} = \tilde\sigma_{+-} (u_- - u_+) ,
\end{equation}
where $\tilde\sigma_{+-}$ is the effective barrier conductance to be determined.

The numbers of particles in the wells, $N_\pm$, calculated according to 
Eq. (\ref{umdef}) are given by
\begin{equation}\label{npm}
N_\pm = N_\pm^{(0)} u_\pm , \qquad N_\pm^{(0)} = Z_\pm /Z ,
\end{equation}
where $Z =Z_+ + Z_-$ is the spin partition function and $Z_\pm$ are
the partition functions in each of the wells.
For the latter it is convenient to introduce the reduced variables
\begin{equation}\label{xial}
\xi \equiv \frac{SH_z}{T}, 
\qquad \alpha \equiv \frac{S^2D}{T},
\qquad h_z \equiv \frac{\xi}{2\alpha} = \frac{H_z}{2SD},
\end{equation}
which are equivalent to those used for the description of classical 
single-domain magnetic particles. \cite{bro63,garishpan90}
Then in the case of not too strong bias $h_z\ll 1$, at low temperatures
the partition functions have the forms
\begin{equation}\label{partfunc}
Z_\pm \cong \frac{ e^{\alpha\pm\xi} }{ 1 - e^{-2\alpha/S} },  
\qquad 
Z \cong \frac{ 2\cosh\xi e^\alpha }{ 1 - e^{-2\alpha/S} } .
\end{equation}
Combining now Eqs. (\ref{nderiv}), (\ref{ohm}), and (\ref{npm}) one comes
to the rate equations
\begin{equation}\label{nderivfin}
\dot N_\pm = \tilde\sigma_{+-} 
\left( 
\frac{ N_\mp }{ N_\mp^{(0)} } - \frac{ N_\pm }{ N_\pm^{(0)} }
\right) .
\end{equation}
For the average spin polarization 
\begin{equation}\label{mzdef}
m_z \equiv \langle S_z \rangle  \cong S(N_+ - N_-)
\end{equation}
the latter result in
\begin{equation}\label{mzeq}
\dot m_z = - \Gamma (m_z - m_z^{(0)}), 
\qquad \Gamma = \frac{ \tilde\sigma_{+-} }{  N_+^{(0)}  N_-^{(0)} } , 
\end{equation}
where, according to Eqs. (\ref{npm}) and (\ref{partfunc}), 
$ N_+^{(0)}  N_-^{(0)} = (4 \cosh^2\xi)^{-1} $.

Finding the effective barrier conductance $\tilde\sigma_{+-}$ determined
by Eq. (\ref{ohm}) is the easiest task in the case without a transverse field
where $\sigma_{mm'}=0$.
Here the elementary resistances $\sigma_{m,m+1}^{-1}$ of
Eq. (\ref{sigmn}) add with the result
\begin{equation}\label{sumres}
\tilde\sigma_{+-}^{-1} = \sum_{m=-S}^{S-1} \sigma_{m,m+1}^{-1} .
\end{equation}
For the thermoactivation rate $\Gamma$ this yields
\begin{equation}\label{gamtha}
\Gamma \cong \frac{4\cosh^2\xi}{Z(\xi,\alpha)}
\left[
\sum_{m=-S}^{S-1} \frac{ \exp(\varepsilon_m/T) }
{\bar l_{m+1,m}^2 W_{m+1,m} }
\right]^{-1} .
\end{equation}
One can see that the main contribution to this expression comes
from the top region, so that $\Gamma \propto \exp[-\alpha(1-h_z)^2]$
and the exact limits of summation in 
Eqs. (\ref{sumres}) and Eq. (\ref{gamtha}) are irrelevant.
Formula (\ref{gamtha}) is the microscopic generalization of the Brown's
result \cite{bro63} on systems with a discrete spectrum.
For $S=1$ a similar result was obtained in early work by Orbach, \cite{orb61} and for a general spin generalizations were given in 
Refs. \onlinecite{vilharsesret94} and \onlinecite{gar97pre} in the
unbiased and biased cases, correspondingly.
In Ref. \onlinecite{gar97pre} different limiting forms of the prefactor
in Eq. (\ref{gamtha}) were analyzed.
The most striking of its features is its dependence on the bias field 
$H_z$ with a strong decrease in the region
where two levels at the top of the barrier come into resonance.
The latter is due to the frequency dependence (\ref{w1res}) of the
one-phonon transition rate between these levels, $W_{m+1,m}$.

In the case of a nonzero transverse field the barrier conductance 
$\tilde\sigma_{+-}$ can be calculated by a well-known recurrence procedure
starting from the top of the barrier.
Introducing $\tilde\sigma_{mm'}$ as the total conductance due to the part of
the barrier between the ``points'' $m$ and $m'$ (see Fig. \ref{t_levs})
one obtains
\begin{equation}\label{recurr}
\tilde\sigma_{mm'} = \sigma_{mm'} + \frac{ 1 }
{
\tilde\sigma_{m+1,m'-1}^{-1} + \sigma_{m,m+1}^{-1} + \sigma_{m',m'-1}^{-1}
}
\end{equation}
with a proper initial condition at the unperturbed  top of the barrier , 
$m_{\rm max}\sim H_z/(2D)$.
If the spin is large and the transverse field $H_x$ is not too small,
the level pair $m_b,m'_b$ corresponding to the actual renormalized
top of the barrier is situated many ``steps'' below $m_{\rm max}$
[see Eq. (\ref{mb})].
In this case the starting point $m_{\rm max}$ becomes unimportant, 
and the recurrence algorithm (\ref{recurr}) generates a continued fraction.
In the Arrhenius regime $T\ll U$, the quantity $\tilde\sigma_{mm'}$
rapidly converges to $\tilde\sigma_{+-}$
down from the renormalized  top of the barrier  $m_b,m'_b$. 
The role of different terms in Eq. (\ref{recurr}) can be made clear if
one considers the ratio
\begin{equation}\label{sigrat}
\renewcommand{\arraystretch}{1.5}
\frac{ \sigma_{mm'} }{ \sigma_{m+1,m} } \sim
\left\{
\begin{array}{ll}
\displaystyle
\frac{ \Omega_{mm'}^2 }{ \omega_{m+1,m}^2 }, 
& |\omega_{mm'}| \sim |\omega_{m+1,m}| ,
\\
\displaystyle
\frac{ \Omega_{mm'}^2 }{ \Gamma_{mm'}^2 }, 
& |\omega_{mm'}| \ll \Gamma_{mm'} ,
\end{array}
\right.
\end{equation}
corresponding to the nonresonant and resonant situations.
If this ratio is of order unity for some pair $m_b,m'_b$, one can consider
all the tunneling conductances $\omega_{mm'}$ above this level as infinite
and below this level as zero [see Eq. (\ref{relspl})].
In the resonant situation, one also can speak about conducting and blocked
resonances.
Since at the level $m_b+1,m'_b-1$ the circuit is completely shunted, one
concludes that renormalized by the transverse field the top of the 
barrier is localized at $m=m_b$, with an uncertainty of one level.
In the nonresonant situation for $1\ll |m| \ll S$ this leads to the
previously obtained {\em classical} result of Eq. (\ref{mb}).
At resonance, for $H_z=0$, the corresponding value of $m_b$ is determined by
the equation
\begin{equation}\label{mbgam}
2S^2 h_x = m_b^2 \left( \frac{ \pi\Gamma_{m_b,m'_b} }{ 2D |m_b| } \right)
^\frac{ 1 }{2|m_b|} .
\end{equation}
Since the level linewidths are small, $\Gamma_{mm'}\ll D$, this value of 
$m_b$ is greater than that off resonance, which thus leads to the resonant
dips in the effective barrier height.
Note, however, that the magnitude of these dips is strongly reduced by the
exponent $1/(2|m_b|)$ in Eq. (\ref{mbgam}), so that they become small
in systems of large spin.  
The shape of resonances in the escape rate $\Gamma$ of Eq. (\ref{mzeq}) can
be visualized, if one considers resonant transitions between  
only one pair of levels $m_b, m'_b$.
Neglecting transitions above this level, one writes
\begin{equation}\label{oneres}
\tilde\sigma_{+-} = \frac{ 1 }
{
\sigma_{-,m_b}^{-1} + \sigma_{m_b,m'_b}^{-1} + \sigma_{m'_b,+}^{-1} ,
}
\end{equation}
where $\sigma_{-,m_b}^{-1}$ is the conductance between the bottom of the 
left well and the point $m_b$, etc.
This expression can be rewritten with the use of Eq. (\ref{sigmn}), and for the
escape rate $\Gamma$ one obtains 
\begin{equation}\label{oneresfin}
\Gamma \cong \frac{ \Omega_{m_b,m'_b}^2 }{ 2 N_+^{(0)}  N_-^{(0)} } \;
\frac{ \Gamma_{m_b,m'_b} N_{m_b}^{(0)} }
{ \omega_{m_b,m'_b}^2 + \Gamma_{m_b,m'_b}^2 + A \Omega_{m_b,m'_b}^2},
\end{equation}
where 
$A=\Gamma_{m_b,m'_b} N_{m_b}^{(0)}( \sigma_{-,m_b}^{-1} + \sigma_{m'_b,+}^{-1} )$.
From Eqs. (\ref{sigmn}) and (\ref{gammm'}) it follows that $A\sim 1$, 
if the resonant transitions through the lower-lying pairs of levels are
neglected.
Thus, contrary to what could be naively expected, the linewidth of the resonance in the escape rate $\Gamma$ is insensitive to the level
linewidth $\Gamma_{m_b,m'_b}$ which is smaller than the tunneling frequency
$\Omega_{m_b,m'_b}$ for conducting resonances.
This frequency grows rapidly with the transverse field.
When it reaches the level spacing $|\omega_{m+1,m}|$, 
the resonance broadens away.
But there are tunneling resonances between  lower pairs of levels
for which the same formula (\ref{oneresfin}) can be written.
The width of these peaks $\Omega_{m,m'}$ is much smaller, but their height
at resonance $\sim \Gamma_{m,m'}N_m^{(0)}$ increases with the level depth as
the Arrhenius factor $N_m^{(0)}\sim \exp(-\varepsilon_m/T)$
and is maximal for the deepest {\em unblocked} pair of resonant levels.
In fact, in the low-damping case the line shape of $\Gamma$ 
described by the continued fraction (\ref{recurr})
consists of many peaks of stepwise decreasing width $\sim\Omega_{m,m'}$ 
mounting on top of each other and forming a self-similar structure.

An illustration of the behavior of the escape rate $\Gamma$ in the
Arrhenius regime based on numerical calculations of the barrier conductance
$\tilde\sigma_{+-}$ will be given in Sec. \ref{numres}.
In the next section we briefly discuss the range of lower temperatures where
a ``more quantum'' behavior of $\Gamma$ is to be expected.

\section{Tunneling versus thermal activation}
\label{tat}

In the Arrhenius regime above, the product $\Omega_{mm'}^2 N_m^{(0)}$ in
the tunneling conductance $\sigma_{mm'}$ of Eq. (\ref{sigmn}) increases
unlimitedly up the barrier and $\sigma_{mm'}$ shunts the effective
circuit at some level $m_b$ determining the renormalized position of the
top of the barrier.
This mechanism is of resonant character, but the temperature dependence
of the escape rate remains classical.
With lowering temperature the question arises, of which group of levels the
tunneling conductance $\sigma_{mm'}$ has a maximum.
The analysis of the function 
$f(m)=\Omega_{mm'}^2 \exp(-\varepsilon_m/T)$ shows
that there are two more regimes in addition to the Arrhenius one --- 
ground-state tunneling and thermally assisted tunneling.  
The temperature of the crossover between these two regimes, $T_{00}$, 
is determined from the condition $f(-S)=f(-S+1)$; i.e., the rate of tunneling
from the first and other excited states falls below the
ground-state tunneling rate.      
The value of  $T_{00}$ calculated with the help of Eq. (\ref{relspl})
has the form
\begin{equation}\label{t00}
T_{00} = \frac{ SD }{ \ln(e^2S/h_x^2) }, \qquad h_x\equiv \frac{ H_x}{2SD} . 
\end{equation}
In theories of tunneling using continuous level models the quantity
$T_{00}$ does not appear.
For models with discrete levels one should keep in mind that the linewidth of the ground states is much smaller than that of excited ones, and
this should make the analysis more subtle, but we will not further pursue
this topic here.

For $T\geq T_{00}$ tunneling goes
through the group of levels between the bottom and the top
for which $f(m)$ has a maximum; if the position of this group 
does not 
coincide with the top of the barrier $m=m_b$, this regime is called 
thermally assisted tunneling.
There are different scenarios for the temperature dependence of this group
of levels, $m\sim m_{\rm TAT}$.
It can shift continuously from the bottom to the top with a crossover to
the Arrhenius regime at some temperature $T_0$.
The other type of behavior is realized if the function $f(m)$ has two 
maxima, say, at the top and near the bottom of the barrier.
In this case there are two competing channels of relaxation which go from
one into the other at the crossover temperature $T_0$.
Both of these scenarios were studied 
for the models with continuous spectra,
and the analogy with the second- and first-order phase transitions was 
pointed out. \cite{chu92}  

\begin{figure}[t]
\unitlength1cm
\begin{picture}(11,7)
\centerline{\epsfig{file=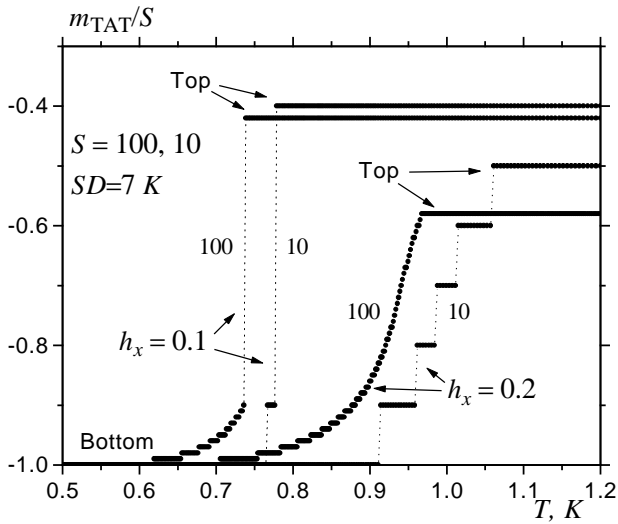,angle=-90,width=12cm}}
\end{picture}
\caption{ \label{t_tat} 
Temperature dependence of the group of levels, $m_{\rm TAT}$, 
making the dominant
contribution into the thermally assisted tunneling, determined from the
maximum of $f(m)=\Omega_{mm'}^2 \exp(-\varepsilon_m/T)$ in the unbiased
case.
}
\end{figure}

For the uniaxial spin model both types of thermally assisted tunneling
can be realized, and the situation can be controlled by the transverse field.
In particular, for the second-order transition the crossover temperature
$T_0$ obtained with the help of Eq. (\ref{relspl}) is given by
\begin{equation}\label{t0}
T_0^{(2)} = SDh_x^{1/2}\frac e8 
\left( 1 - \frac{e^2}{8} \frac{h_x}{2} \right). 
\end{equation}
For low transverse fields $T_0^{(2)}$ becomes too small, and the first-order
transition to the regime of thermally assisted tunneling occurs when 
the temperature is lowered before $T_0^{(2)}$ is reached.
Details of the analysis will be presented elsewhere; here we illustrate
the temperature dependence of $m\sim m_{\rm TAT}$ in Fig. \ref{t_tat}.
It can be seen that the higher values of $h_x$ favor the second-order transition: 
The curve $m_{\rm TAT}(T)$ goes ``continuously''
through each value of $m$ and merges at $T_0$ with the horizontal line
$m=m_b$ characterizing the Arrhenius regime.
On the contrary, in lower fields $h_x$ large jumps of $m_{\rm TAT}$ at $T_0$
can be seen. 
For smaller spins the low-temperature tail of the curve $m_{\rm TAT}(T)$ 
becomes shorter.
The value of $T_{00}$ is in all cases well described by formula 
(\ref{t00}).

In the thermally assisted tunneling regime, the ratio of the tunneling 
and intrawell conductances, Eq.
(\ref{sigrat}) is a very small number in the relevant region 
$m\sim m_{\rm TAT}$.
Thus the slow tunneling process controls the escape rate $\Gamma$, and the
distribution of particles in the wells does not deviate from 
quasiequilibrium.
In this case $\Gamma$ is simply given by
\begin{equation}\label{gamlt}
\Gamma = \frac{ \tilde\sigma_{+-} }{  N_+^{(0)}  N_-^{(0)} } , 
\qquad 
\tilde\sigma_{+-} = \sum_{m=-S}^{m_{\rm max}} \sigma_{mm'} ,
\end{equation}
i.e., it is the tunneling probability weighed with the Boltzmann factor
[see Eq. (\ref{sigmn})].   
Expressions similar to 
Eq. (\ref{gamlt}) were taken as a starting point in many investigations    
of the escape rate of particles from a metastable well at nonzero 
temperatures (see, e.g., Ref. \onlinecite{aff81}).
An efficient method of treating this problem for {\em continuous}
spectra, including the dissipative case, is based on the instanton
technique. \cite{calleg83,larovc83}
For our spin model, however, the spectrum cannot be made continuous by
a reasonable
variation of some physical parameter; the tunneling frequency changes 
abruptly from one level to another, and this situation persists in the
limit $S\to \infty$ (see the end of Sec. \ref{qmech}). 
This situation seems to be pertinent not only to spin systems, which can be,
in fact, mapped onto the particles, \cite{schwrelvh87,zas90} but for 
double-well models in general.
Resonant tunneling between the discrete levels in a 
low-damped SQUID was observed recently in Ref. \onlinecite{rousiyluk95}.
The numerically calculated tunneling level splittings for the SQUID 
Hamiltonian \cite{rousiyluk95} also change abruptly
from one level pair to another.

The advantage of our more general approach to finding the barrier
conductance $\sigma_{+-}$ based on the recurrence relations
(\ref{recurr}) in comparison to the simplified formula (\ref{gamlt}) is
its ability to handle the case of very small coupling to the bath.
In this case the relaxation rates for the exchange between the
neighboring levels $\sigma_{m,m+1}$ of Eq. (\ref{sigmn}) become very
small, as well as the tunneling conductances  $\sigma_{mm'}$ off
resonance, and so does the resulting escape rate $\Gamma$.
If one sets the system on resonance to increase tunneling, then
the system does not come to quasiequilibrium in each of the wells and
formula  (\ref{gamlt}) breaks down.

\section{Numerical results for the escape rate; role of nuclear spins
and the axis misalignment}
\label{numres}

In this section we present the results of numerical simulations for the
escape rate $\Gamma$ obtained with the methods of the previous section
in the Arrhenius regime.
The region below the crossover temperature $T_0$ is
not further considered in this paper.
For systems of moderate spin the range of thermally assisted tunneling 
is rather narrow, and at temperatures $T\leq T_{00}$ in the unbiased case
tunneling should go between the ground states.
Since the linewidths of the ground-state levels are exponentially small
at such temperatures, even a small detuning is sufficient to suppress
the resonance.
In this case we face a strongly nonresonant situation, and our
theoretical methods of Sec. \ref{slowdyn} should be modified.
Even in the Arrhenius regime, there is a problem with nonresonant
processes --- the escape rate calculated with the help of 
Eq. (\ref{recurr}) shows discontinuities at values of the bias field,
at which we switch from one resonant partner level to another in the 
calculation routine.
In fact, the resonance of each level with {\em several} partners in the
other well should be considered, but a rigorous treatment of this
problem would lead to serious complications.
For this reason we simply extend the applicability of the kinetic equation (\ref{mastereq})
by considering, for each level $m$, the tunneling resonances with
the two partners $m'$ and $m'-1$ satisfying 
$\varepsilon_{m'} < \varepsilon_{m} < \varepsilon_{m'-1}$.
In this symmetric approach the switching between partners occurs in
resonance and no discontinuity in $\Gamma$ appears.
The calculations in this case can be performed with the help of the modification of the recurrence relation (\ref{recurr}).
The resonance of one level with all other partners was
considered  by Garg \cite{garg95diss} using the damped Schr\"odinger equation.

Treating the relaxation terms we replace $\bar l_{m,m+1}$ by 
$l_{m,m+1}$ [see Eq. (\ref{rmn})], which amounts to dropping the operator
$S_z$ in Eq. (\ref{sphamaa}). 
Then we fit the strength of the spin-phonon coupling
to the measured for Mn$_{12}$Ac value of the prefactor 
$\Gamma_0 = 5\times 10^6 {\rm s}^{-1}$ in the escape rate
$\Gamma = \Gamma_0 \exp(-U_{\rm eff}/T)$.
Since the experimental temperatures about several kelvin exceed the level
spacing near the top of the barier, $\omega_{m+1,m}\sim 2Dm \sim 1$ K,
the prefactor depends linearly on temperature; see the first line
of Eq. (\ref{w1res}).
This dependence is, however, difficult to see in the limited
temperature interval.   

\begin{figure}[t]
\unitlength1cm
\begin{picture}(11,7)
\centerline{\epsfig{file=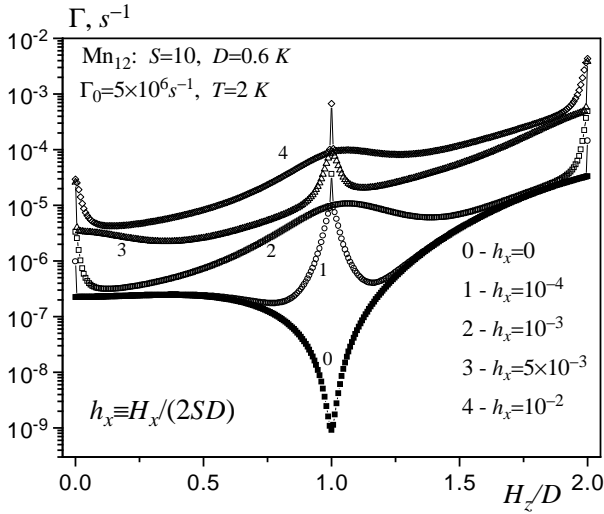,angle=-90,width=12cm}}
\end{picture}
\caption{ \label{t_hz} 
Bias-field dependence of the resonant tunneling escape rate of the
uniaxial spin model.
}
\end{figure}

The results for the escape rate as a function of the bias field $H_z$ are 
represented on Fig. \ref{t_hz} for different values of the transverse field $H_x$.
One can see the superpositions of broad and narrow peaks at the resonant
values of the bias field $H_{zk}=Dk$, which correspond to the tunneling
via the shallower and deeper resonant levels, respectively.
The width of peaks alternates as a function of the resonance number $k$, 
since the tunneling transitions between different pairs of levels appear
in even or odd orders of the perturbation theory in $H_x/D$; 
see Sec. \ref{qmech}.
In particular, in the unbiased case $k=0$, the tunneling between the 
level pair $-1,1$ appears in second order, 
$\Omega_{-1,1}\propto H_x^2/D$.
As a result, a very narrow peak in $\Gamma$ emerges at $H_z=0$ for 
$h_x=10^{-4}$.
This peak broadens with the increase of $H_x$, and at 
$h_x=5\times 10^{-3}$ a new narrow peak corresponding to the resonance
$-2,2$ with $\Omega_{-2,2}\propto H_x^4/D^3$ is seen.
A similar picture holds for $k=2$ and other even resonances.

For the odd resonances, as $k=1$, the escape rate in zero transverse field becomes small due to the frequency dependence of one-phonon
processes discussed above.
The same result was obtained for the tunneling assisted one-phonon processes
between the deep levels in the wells. \cite{polretharvil95}
In Fig. \ref{t_hz} we have included a small frequency-independent 
contribution from the Raman scattering processes to obtain a nonzero
value of the escape rate. \cite{gar97pre}
This feature is, however, completely suppressed already in very small
transverse fields because of the opening of a new transition channel:
the tunneling between the topmost resonant pair $-1,0$ that appears  
in the first order, $\Omega_{-1,0}\propto H_x$.
The latter is, in fact, a kind of a free precession around ${\bf H}_x$ rather
than tunneling. 
The rate of this precession competes with the small relaxation rate 
[see the second line of Eq. (\ref{sigrat})]; that is, 
the purely dynamical transition between the levels $-1,0$ competes 
with the dissipative one.
As a result, the dip in $\Gamma$ yields to the massive peak already
for $h_x=10^{-4}$ for the damping parameters appropriate for Mn$_{12}$Ac.

For higher values of the transverse field $h_x$, the behavior of the even and
odd resonance peaks is the same.
As $h_x$ is growing, the condition 
$\Omega_{mm'}(h_x)\geq \Gamma_{mm'} $ of Eq. (\ref{sigrat})
for a given pair of levels $m,m'$
is satisfied at a certain value of the transverse field $h_{xb}$.
At that value the $m,m'$ resonance becomes unblocked.
This results in a new narrow peak, of 
width about $\Omega_{mm'}(h_x)$, which appears on the top of the 
$m+1,m'-1$ resonant peak (see Fig \ref{t_hz}).
This situation is quite universal in the sense that the $m+1,m'-1$ resonant peak
can itself be a narrow peak on the top of the $m+2,m'-2$ resonant peak.
In general, each resonance consists of a few peaks mounting on top of
each other.
The width of two consequent peaks within one resonance differs by a factor about $e^4\sim 55$, in accordance with Eq. (\ref{relspl}). 
The magnification of the $H_z$ interval around the resonant values $H_{zk}$ shows the self-similar multitower structure of the resonance.
In that structure the total number of peaks depends
on the strength of the dissipation, while their height is determined by
temperature.
The lower the damping, the greater is the number of the peaks.
The lower the temperature, the greater is the difference in the height of the
peaks mounting on top of each other.     

\begin{figure}[t]
\unitlength1cm
\begin{picture}(11,7)
\centerline{\epsfig{file=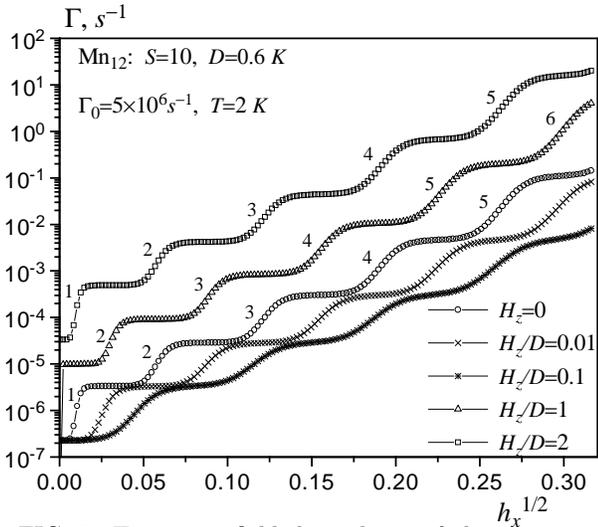,angle=-90,width=12cm}}
\end{picture}
\caption{ \label{t_hx_sq} 
Transverse-field dependence of the resonant tunneling escape rate of the
uniaxial spin model.
}
\end{figure}

\begin{figure}[t]
\unitlength1cm
\begin{picture}(11,8)
\centerline{\epsfig{file=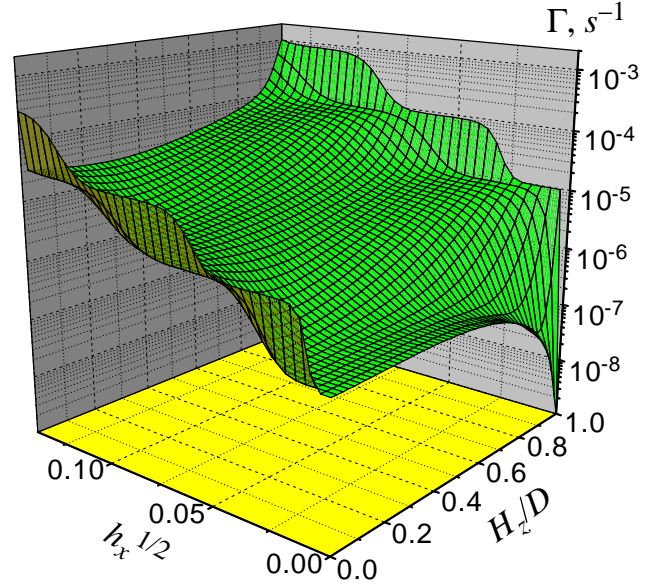,angle=-90,width=15cm}}
\end{picture}
\caption{ \label{t_3d2} 
Dependence of the escape rate of the
uniaxial spin model on the field components $H_x$ and $H_z$
(parameters are appropriate for Mn$_{12}$Ac).
}
\end{figure}

The dependences of $\Gamma$ on the transverse field $H_x$ for the 
resonant and slightly off-resonance values of the bias field are shown
in Fig. \ref{t_hx_sq}.
The steps on the resonant $H_x$ dependences of $\Gamma$ correspond to
the values of $H_x$ at which the value of $m_b$ determined from 
Eq. (\ref{mbgam}) takes an integer value (or a half-integer value
for systems of half-integer spin $S$).
For these values of $H_x$ a resonant shunting of the barrier at the
next deeper level occurs.
The flat regions correspond to the situation when one pair of resonant
levels is already completely shunted and the following (the lower) one is
yet completely unshunted.
The step values of $H_x$ are sensitive to the sum of the level linewidths
$\Gamma_{m_b,m'_b}$ given by Eq. (\ref{gammm'}), and thus such 
experiments are conceivable as a kind of spectroscopy measuring the 
relaxation characteristics of {\em separate} levels.
The 3$d$ plot of $\Gamma(H_x,H_z)$ summarizing the features of 
resonant tunneling process discussed above is presented in 
Fig. \ref{t_3d2}.

\begin{figure}[t]
\unitlength1cm
\begin{picture}(11,7)
\centerline{\epsfig{file=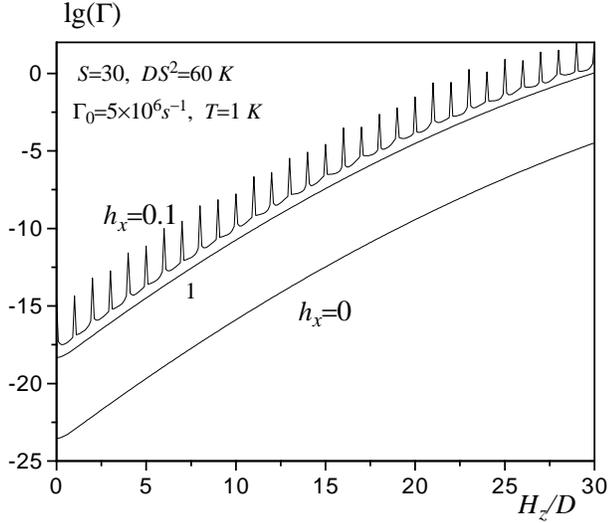,angle=-90,width=12cm}}
\end{picture}
\caption{ \label{t_class} 
Longitudinal field dependence of the escape rate of a large-spin 
uniaxial model in the fixed transverse field.
The curve 1 was obtained from the pure thermoactivation curve $h_x=0$
using Eq. (\protect\ref{classbarlowxz}) for the barrier lowering in the 
transverse field $h_x=0.1$.  
}
\end{figure}

Apart from resonant tunneling, the overall shape of $\Gamma(H_x,H_z)$
follows approximately the Arrhenius law with the classical barrier height  
$U(H_x,H_z)$.
This can be seen especially clear for systems with large spin, 
frequency-independent relaxation rates and low 
temperatures.
The last condition is needed to reduce the relative role of the
field dependence of the prefactor $\Gamma_0 = \Gamma_0(H_x)$ 
in the classical expression
for $\Gamma$, which is not yet well established
(see Refs. \onlinecite{bro79} and \onlinecite{cofetal95}). 
The comparison of our calculation with the classical result accounting
only for the dependence 
\begin{equation}\label{classbarlowxz}
U(H_x,H_z) \cong DS^2(1-h_z)^2 
\left[1 - 2h_x \frac{(1-h_z^2)^{1/2}} {(1-h_z)^2 } \right]
\end{equation}
for $h_x\ll 1$ is presented in Fig. \ref{t_class}.
The rather good accordance between the classical and quantum results
illustrates the conjectures of Sec. \ref{qmech} in a more general 
biased case.

The resonant tunneling curves obtained above do not fully explain
the experimental observations 
\cite{frisartejzio96,heretal96,thoetal96,heretal97,lioetal97} 
showing that all peaks have approximately the same form.
The latter can be the consequence of the averaging effect due to the 
misalignment of the particle's axes in not perfectly oriented
polycrystalline samples.
A similar effect can be caused in Mn$_{12}$Ac by nuclear spins
whose fluctuating transverse components can, in addition, induce
tunneling even in the absense of an externally applied field $H_x$.
The corresponding adjustments of our method will be made below.

\begin{figure}[t]
\unitlength1cm
\begin{picture}(11,7)
\centerline{\epsfig{file=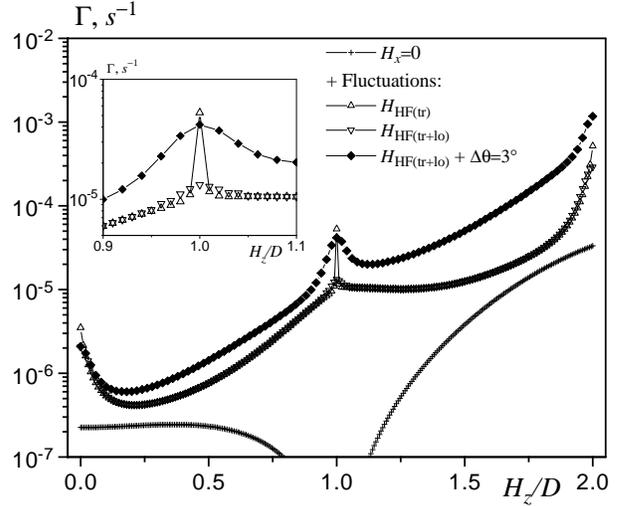,angle=-90,width=12cm}}
\end{picture}
\caption{ \label{t_fl} 
Resonant tunneling escape rate in Mn$_{12}$Ac due to nuclear spins and
the crystallite misalignments.
}
\end{figure}

In a Mn$_{12}$Ac molecule each of 12 Mn atoms interacts with its own
nuclear spin ${\bf I}_i$, $I=5/2$, via the hyperfine (HF) interaction.
For the total cluster spin this interaction can be approximately written
as
\begin{equation}\label{hfham}
{\cal H}_{\rm HF} \simeq -A_{\rm eff} {\bf S} {\bf I}_{\rm tot} ,
\qquad  {\bf I}_{\rm tot}\equiv \sum_i {\bf I}_i .
\end{equation}
In fact, the hyperfine interactions are somewhat different for
different Mn atoms, and their extensive discussion can be found in 
Ref. \onlinecite{harpolvil96}.
If all the nuclear spins are aligned in the same direction, the energy
of the HF interaction $E_{\rm HF, max}\simeq 12ISA_{\rm eff}\simeq 0.6$ K
is comparable to the level spacing near the top of the barrier 
$|\omega_{m+1,m}|\sim 2Dm \sim 1$ K and is much greater than the 
dipole-dipole energy $E_d\simeq 0.06$ K. 
The effective HF field produced 
by the nuclei on the cluster spin is in this case about
$H_{\rm HF, max}\simeq E_{\rm HF, max}/(g\mu_BS) \simeq 0.05$ T.
If this HF field is perpendicular to the easy axis $z$, the corresponding
dimensionless transverse field 
$h_{x, {\rm HF}} = g\mu_BH_{\rm HF, max}/(2SD) \simeq 3.7\times 10^{-3}$
should result in strong resonant (as well as nonresonant) tunneling;
see Fig. \ref{t_hz}.
On the other hand, the role of the $z$ component of the HF field in  
resonant tunneling is determined by the dimensionless parameter
$g\mu_BH_{\rm HF, max}/D \simeq 0.075$.
This shows that the narrow 
resonance lines in Fig. \ref{t_hz} should be averaged away by the
fluctuating $z$ component of the hyperfine field; i.e., the hyperfine
interaction suppresses the coherence.
This second effect was discussed by several authors; 
\cite{garg95nuccoh} here we will take into account both effects of 
nuclear spins with the help of simplified qualitative arguments.

The subtlety of the hyperfine interaction is that it conserves the total
projection $S_z + \sum_i I_{iz}$, and, strictly speaking, the coupled
equations of motion for the tunneling cluster spin and rotating nuclear
spins should be solved.
In the Arrhenius regime, however, tunneling occurs near the top of the
barrier where it is rather fast --- it ranges from 
$\Omega_{mm'}\sim |\omega_{m+1,m}|$ off
resonance to $\Omega_{mm'}\sim \Gamma_{mm'}$ at resonance.
This is much faster than the nuclear relaxation rate which is due to the 
fluctuating magnetic fields and is determined by the small nuclear
magnetic moment.
Further, tunneling of the cluster spin near the  top of the barrier  leads to a
relatively small change of its $z$-projection: 
$\Delta S_z\simeq 2m_b \ll 2S$.
This is not a large part of the whole integral of motion
$S_z + \sum_i I_{iz}$.
Indeed, for the randomly oriented nuclear spins the second term of this
sum is on average of order $\sqrt{12}I \simeq 8.7$, and thus tunneling of the
cluster spin can be compensated by the corresponding rotation of the
nuclear spins.
(On the contrary, for tunneling from the ground state at $T\leq T_0$
the $z$ projection change is 
$\Delta S_z = 2S =20$, and this process cannot go via the interaction
with the nuclear spins --- it is blocked by the
conservation law.)   
Thus, in the Arrhenius regime one can qualitatively consider nuclear
spins as frozen --- they do not change their state as a result of the
tunneling of the cluster spin.
The distribution function of the HF field on the cluster spin can be easily found.
As the energy of the interaction of {\em one} nuclear spin with the
cluster spin $ISA_{\rm eff}\simeq 0.05$ K is much smaller than
temperature, one can use the infinite-temperature distribution function
for the individual nuclear spins.
Then, for a large number of nuclear spins, $N=12\gg 1$, the 
quantum-statistical averages of the total nuclear spin 
${\bf I}_{\rm tot}$
in Eq. (\ref{hfham}) are given by the Gaussian distribution function
\begin{equation}\label{nucdf}
F( {\bf I}_{\rm tot} ) = \frac{ 1 }{ (2\pi\sigma_I)^{3/2} }
\exp\left( - \frac{ I_{\rm tot}^2 }{ 2\sigma_I } \right),
\end{equation}
where the dispersion  $\sigma_I=(N/3) I(I+1)$ can be checked calculating
the average $\langle I_{{\rm tot},z}^2\rangle$ directly and from
Eq. (\ref{nucdf}) and comparing the results.
Now, all the previously obtained expressions for the escape rate $\Gamma$,
as well as such quantities as the time dependence of magnetization and
dynamic susceptibility, should be averaged with the distribution function
$F$.
In the absense of an externally applied field $H_x$, the averaging of each
quantity $ A(H_x,H_z) $ is done explicitly as
\begin{eqnarray}\label{avr}
&&
\bar A(H_x,H_z) = \!\int\limits_{0}^{\infty}\!\! dx\;2x e^{-x^2}\!\!\!
\int\limits_{-\infty}^{\infty}\!\!\! dz\; \frac{ e^{-z^2} }{ \pi^{1/2} }
A(x\bar H_{\rm HF},H_z\!+\!z\bar H_{\rm HF}) ,
\nonumber\\
&&
\bar H_{\rm HF} = H_{\rm HF, max} \frac{(2\sigma_I)^{1/2} }{ NI},
\qquad
H_{\rm HF, max} = \frac{ NIA_{\rm eff} }{ g\mu_B }.
\end{eqnarray}
The results of this averaging for the escape rate $\Gamma$ are presented
in Fig. \ref{t_fl}.
The role of nuclear spins in inducing 
resonant tunneling and suppressing the narrow resonance lines is clearly
seen.
In addition, we have taken into account small fluctuations of the
directions of the anisotropy axes of Mn$_{12}$Ac molecules in 
polycrystalline samples with the dispersion of only $\Delta\theta=3^\circ$.
These misalignments also produce a static fluctuating components of the
transverse field, and their role becomes progressively more 
important with the increase of the bias field $H_z$.
One can see that when all these effects are taken into account, all the
resonant tunneling peaks become approximately of the same form, as
observed in experiments.

\section{Discussion}
\label{disc}

We have presented the theory of thermally activated resonant spin tunneling.
The bulk of the theory applies to any molecular magnet, while particular 
numerical illustrations were made for Mn$_{12}$Ac.
Quantization of spin levels, which is the key to explaining experimental
results, has dictated our choice of theoretical apparatus.
Rather than employing instanton methods, suitable for models with continuous 
spectra, we have used the density matrix description of the spin interacting with
thermal bath.

In continuous models three regimes for the escape rate $\Gamma$ are usually
studied.
At high temperatures quantum-mechanical effects are not important, and the 
escape over the barrier is due to pure thermal activation described by the
Arrhenius law.
In the limit of zero temperature only tunneling out of the ground state is 
important.
There is also an intermediate regime which combines thermal activation to
excited levels with tunneling across the barrier, which is called 
thermally assisted tunneling. 
In that regime the position of the narrow group of levels which dominate the
escape rate depends on temperature, moving continuously from the ground state
at $T=0$ to the top of the barrier at the temperature called the crossover temperature.
This situation describes a conventional, smooth, second-order transition from 
quantum tunneling to thermal activation. \cite{aff81}
In principle that transition can also be first order, which would
correspond
to the sharp crossover from quantum tunneling to thermal
Arrhenius-type behavior. \cite{chu92} 
We have demonstrated that this is exactly what happens for a spin system
in low transverse field.
Correspondingly, the experimental study of the escape rate should find the
evolution from sharp to smooth crossover between thermally assisted tunneling 
and the Arrhenius regime on the transverse field.

In systems of moderate spin, such as Mn$_{12}$Ac, thermally assisted tunneling
occurs in a rather narrow temperature range.
In experiments the Arrhenius law that occurs in a wider temperature range
has been observed.
Despite the purely classical temperature dependence of the relaxation
in the Arrhenius regime,
the field dependence of $\Gamma$ shows quantum effects due to the discrete
nature of spin ignored in continuous models.
Contrary to these models, which start with a given barrier, a well-defined
barrier does not exist for a mesoscopic spin;
its effective value depends on the bias field $H_z$ in a nonmonotonic
manner.
The observed minima of the effective barrier are due to the crossing of the
spin levels, which results in resonant tunneling between the wells. 
This is different from a classical spin system where the barrier monotonically
decreases with increasing $H_z$.
This regime can be called {\em thermally activated tunneling}, 
as different from the regime of thermally assisted tunneling. 
The difference between the two regimes is that in the first regime tunneling
always occurs at the top of the barrier, while in the second regime it
occurs from excited levels between the bottom and the top of the barrier.

The theory predicts that each resonance in the escape rate $\Gamma$ has a 
multitower structure with peaks of decreasing width mounting on top of 
each other.
This effect is due to resonant spin tunneling between different matching 
levels.
All peaks are centered at the same field, if the corresponding pair of levels
match at the same value of the bias field.
Note that this assumpion relies on the simple form of the Hamiltonian used 
in our calculations.
Additional terms of different symmetry would violate this assumtion.
If these terms are small, as they are in Mn$_{12}$Ac, the resonances on $H_z$ will not be exactly equidistant and the centers of peaks towering in each resonance must be slightly displaced with respect to each other.
The number of peaks in each resonance increases with decreasing dissipation.

Depending on the number $k$ of the resonance [see Eq. (\ref{hzres})], 
the leading
contribution to the rate appears in even or odd orders of the perturbation
theory on the transverse field $H_x$. 
This results in the alternation of the shape of resonances on $H_z$.
Another effect predicted by the theory is the stepwise dependence of the rate
on the transverse field when the longitudinal field is tuned to the resonance.

The origin of the terms in the Hamiltonian responsible for tunneling is 
different for different molecular magnets.
The absense of any selection rules for resonances in Mn$_{12}$Ac 
unambiguously points to transverse fields causing the transitions.
These fields originate from the hyperfine and, to a smaller degree, from the 
dipole-dipole interactions.
In Fe$_8$ the hyperfine interactions are negligible, and the transitions are
presumably caused by the transverse anisotropy.

Our theory for Mn$_{12}$Ac can pretend to the quantitative description of the
magnetic relaxation in this system, as it takes into account all major
contributions to the effect.
However, observation of more subtle effects, such as the multitower
structure of resonances, the alternating shape, stepwise dependence on the 
transverse field, etc., is less likely in Mn$_{12}$Ac.
This is because of the smearing of these effects by strong fluctuations of the
hyperfine field.
Fe$_8$ (see, e.g., Ref. \onlinecite{sanetal97}) 
seems to be a better candidate for observing these effects.

\section*{Acknowledgments}

We would like to acknowledge support from 
the U.S. National Science Foundation through
Grant No. DMR-9024250.
The work of D.G. has not been supported by the 
Deutsche Forschungsgemeinschaft.

\renewcommand{\thefootnote}{\fnsymbol{footnote}}
\footnotetext[1]{ 
Permanent address: 
I. Institut f\"ur Theoretische Physik, Universit\"at Hamburg,
Jungiusstrasse 9, D-20355 Hamburg, Germany.
Electronic addresses: 
garanin@physnet.uni-hamburg.de,
garanin@t-online.de }

\renewcommand{\thefootnote}{\fnsymbol{footnote}}
\footnotetext[2]{
Electronic address:
chudnov@lcvax.lehman.cuny.edu}


\end{document}